\documentclass[cits,hyper]{JINST}
\usepackage{epsfig,paralist,subfig,multirow}

\newcommand{\pho}{\phantom{0}}
\newcommand{\phm}{\phantom{-}}

\newcommand\T{\rule{0pt}{2.4ex}}        
\newcommand\B{\rule[-0.0ex]{0pt}{0pt}}  

\title{ATLAS silicon module assembly and qualification tests at IFIC Valencia}

\author{J.~Bernabeu, J.~V.~Civera, M.~J.~Costa, C.~Escobar, J.~Fuster,
C.~Garc\'{\i}a, J.~E.~Garc\'{\i}a-Navarro, F.~Gonz\'{a}lez,
S.~Gonz\'{a}lez-Sevilla, C.~Lacasta, G.~Llos\'{a}, S.~Marti-Garc\'{\i}a,
M.~Mi\~{n}ano, V.~A.~Mitsou\thanks{Corresponding author.}, P.~Modesto,
J.~N\'{a}cher, R.~Rodr\'{\i}guez-Oliete, F.~J.~S\'{a}nchez, L.~Sospedra,
V.~Strachko \\
Instituto de F\'{\i}sica Corpuscular (IFIC), CSIC - Universitat de Val\`{e}ncia \\
Edificio Institutos de Investigaci\'{o}n \\
Apartado de Correos 22085, E-46071 Valencia, Spain \\
E-mail: \email{vasiliki.mitsou@ific.uv.es}}%

\abstract{ATLAS experiment, designed to probe the interactions of particles
emerging out of proton proton collisions at energies of up to 14~TeV, will
assume operation at the Large Hadron Collider (LHC) at CERN in 2007. This paper
discusses the assembly and the quality control tests of forward detector
modules for the ATLAS silicon microstrip detector assembled at the Instituto de
F\'{\i}sica Corpuscular (IFIC) in Valencia. The construction and testing
procedures are outlined and the laboratory equipment is briefly described.
Emphasis is given on the module quality achieved in terms of mechanical and
electrical stability.}


\begin{document}
\section{Introduction}\label{sec:intro}

The ATLAS detector \cite{atlas}, one of the two general-purpose experiments of
the Large Hadron Collider (LHC), has entered into the final stages of
installation at CERN. The LHC, a proton-proton collider with a 14-TeV
centre-of-mass energy and a design luminosity of \mbox{$10^{34}~{\rm
cm^{-2}s^{-1}}$,} is expected to deliver the first proton beam by the end of
2007. The ATLAS central tracker (Inner Detector, ID) \cite{TDR} combines the
silicon detector technology (pixels \cite{pixel} and micro-strips \cite{sct})
in the innermost part with a straw drift detector with transition radiation
detection capabilities (Transition Radiation Tracker, TRT) \cite{TRT} in the
outside, operating in a 2-T superconducting solenoid.

\begin{figure}[ht]
    \centering
    \epsfig{file=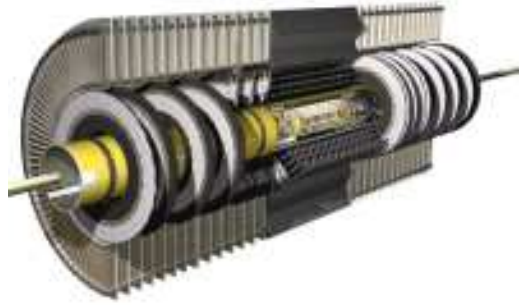,width=0.45\linewidth,clip=}
    \caption{Layout of the ATLAS Inner Detector: it comprises the Transition
    Radiation Detector, the Semiconductor Tracker and the Pixel system
    from the outer to the inner radii, respectively.} \label{fig:ID}
\end{figure}

The microstrip detector (Semiconductor Tracker, SCT) \cite{sct}, as shown in
Fig.~\ref{fig:ID}, forms the middle layer of the ID between the Pixel detector
\cite{pixel} and the TRT \cite{TRT}. The SCT system comprises a barrel made of
four nested cylinders and two end-caps of nine disks each. The barrel layers
carry 2112 detector units {\em (modules)} altogether, while a total of 1976
end-cap modules are mounted on the disks. The whole SCT occupies a cylinder of
5.6~m in length and 56~cm in radius with the innermost layer at a radius of
27~cm.

The silicon modules \cite{barrel,endcap} consist of one or two pairs of
single-sided p-\emph{in}-n microstrip sensors \cite{sensor} glued back-to-back
at a 40-mrad stereo angle to provide two-dimensional track reconstruction. The
$285$-$\mu{\rm m}$ thick sensors have 768 AC-coupled strips with an $80~\mu{\rm
m}$ pitch for the barrel and a $57-94~\mu{\rm m}$ pitch for the end-cap
modules. Between the sensor pairs there is a highly thermally conductive
baseboard. Barrel modules follow one common design, while for the forward ones
four different types exist according to their position in the detector.

\begin{figure}[ht]
\centering
\begin{minipage}[c]{0.43\linewidth}
  \centering\epsfig{file=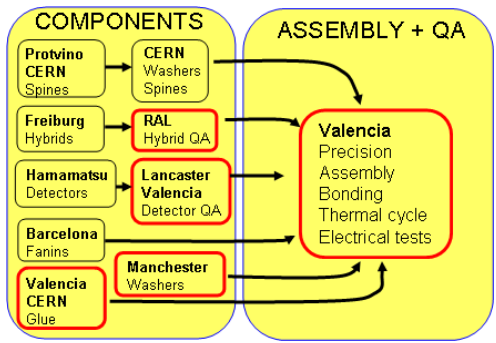,width=\linewidth}
\end{minipage} \hspace{0.05\linewidth} 
\begin{minipage}[c]{0.37\linewidth}
  \centering
  \caption{Flow of components for module production and
  list of assembly tasks and quality control tests performed in Valencia.} \label{fig:flow}
\end{minipage}
\end{figure}

The SCT construction involved ---among other development projects and
macro-assembly--- the building and Quality Control (QC) of \mbox{$\sim4000$}
modules. In particular, the assembly of the required 1976~ATLAS SCT forward
modules \cite{endcap} plus a contingency of 20\% was distributed among
13~European and one Australian institutes, divided into three clusters in order
to facilitate the sharing of tasks and the flow of components. The IFIC group,
as a member of the UK-V cluster, followed the complete assembly sequence and
all the necessary QC tests were performed, as shown in Fig.~\ref{fig:flow}, for
all assembled modules. The full production was launched after a qualification
period, during which the tooling and the execution of the procedures were
successfully validated.

The IFIC-SCT group assembled and tested a total of 282~forward SCT modules;
125~outer modules and 157~long middle ones. Before the production startup, the
Valencia group initially undertook the construction of 125~outer and 96~middle
modules. After a re-organisation of the forward modules production, the group
accepted the responsibility to assemble a surplus of 61~middle modules, which
amounts to a 28\% increase with respect to the initial commitment.

This paper begins with a brief description of the module assembly and QC
sequence in Sec.~\ref{sec:assembly}, followed by a report on the metrology
procedure and respective measurements in Sec.~\ref{sec:metrology}. In
Sec.~\ref{sec:electrical}, the front-end electronics and the electrical setup
and tests are outlined, and the corresponding results are reported. Some
studies performed on modules assembled in other sites are discussed in
Sec.~\ref{sec:CiS}. An overview of the QC tests results and the conclusions are
given in Secs.~\ref{sec:results} and~\ref{sec:conclusions}, respectively.

\section{Module assembly and quality control overview} \label{sec:assembly}

\begin{figure}[ht]
  \centering
  \epsfig{file=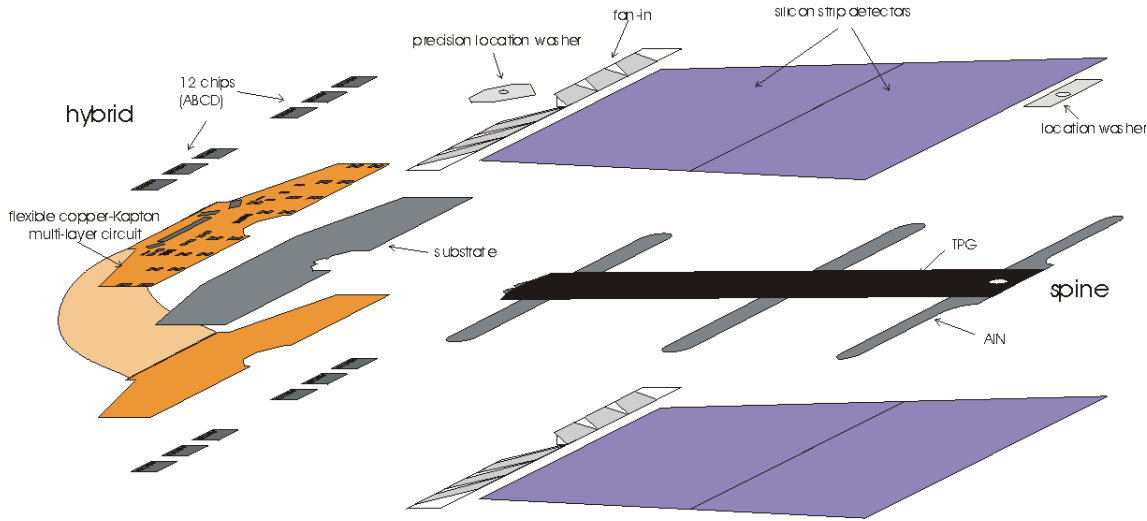,width=\linewidth}
  \caption{The module assembly scheme.}\label{fig:outer_module}
\end{figure}

The forward silicon modules \cite{endcap} consist of one or two pairs of
single-sided p-\emph{in}-n micro-strip sensors glued back-to-back at a 40-mrad
stereo angle, as shown in Fig.~\ref{fig:outer_module}, to provide
two-dimensional track reconstruction. The 285-$\mu{\rm m}$ thick sensors
\cite{sensor,components} have 768~AC-coupled strips implanted with a pitch of
$57-94~\mu{\rm m}$. Between the sensor pairs there is a highly thermally
conductive baseboard \emph{(spine)\/} \cite{components}. The sensors are
connected to the front-end electronics board \emph{(hybrid)\/} \cite{hybrid} by
means of \emph{fan-ins}\footnote{Parts made of glass with aluminum traces,
providing electrical connection channel-by-channel from the sensors to the
read-out chips and mechanical connection between the hybrid and the detector
part of the module.} \cite{components}. Barrel modules \cite{barrel} follow one
common design, while for the forward ones four different types exist based on
their position in the detector. At IFIC, outer and long-middle modules were
built made up out of two pairs of Hamamatsu\footnote{Hamamatsu Photonics Co.\
Ltd., 1126-1 Ichino-cho, Hamamatsu, Shizuoka 431-3196, Japan.} silicon sensors.

In order to carry out the module production according to the quality criteria
set by the ATLAS SCT collaboration, dedicated machinery and tooling was
constructed and installed in a clean room of \mbox{$\sim\rm75~m^2$} at IFIC,
where temperature was controlled within $\rm0.5~^{\circ}C$ and relative
humidity within 5\% (keeping it below 50\%). This area was divided into two
compartments: a small one ($\rm20~m^2$) of class 1000 and a big room
($\rm55~m^2$) of class 10\,000. In the former the electrical properties of the
sensors and modules were measured and the visual inspection of the components
was performed, while in the latter the remaining assembly and testing tasks
took place.

Besides the production-oriented infrastructure, a laser test bench has been
developed aiming at various studies on silicon detectors
---including SCT modules. Specifically, measurements were carried out
in order to detect defects on strips \cite{laser}. Important characteristics of
module performance were determined with the laser beam, such as the response
dependence on interstrip position, the pulse shape reconstruction for different
impact positions along the strip, the spatial resolution, etc.

The end-cap module comprises several components \cite{components}, viz.\ the
silicon sensors, the spine, the fan-ins, the washers and the electronics hybrid
\cite{hybrid}, assembled as shown in Fig.~\ref{fig:outer_module}. Tests were
made upon reception and during assembly to assure that the components have not
been damaged during transport or in the first stages of module construction,
including a first $IV$ scan of the individual silicon sensors. In the
following, the module assembly steps and intermediate quality checks, described
in full detail in Refs.~\cite{endcap,FDR,assembly,FDRQA}, are outlined and only
the technical features specific to the IFIC laboratory are mentioned.

\begin{asparadesc}

\item[Sensor pair alignment.]
To achieve good tracking performance stringent requirements on the positioning
of the modules are imposed. The precise alignment \cite{assembly} of the
detectors is achieved by employing a robot consisting of six linear
stages,\footnote{Nappless-Coombe Ltd.} pictured in Fig.~\ref{fig:assembly:a}.
It is equipped with an optical system for pattern recognition and stages which
are commanded by the DMC-1500 controller\footnote{Galil Motion Control, Inc.}
through a PC using a program running under LabVIEW developed at the University
of Manchester \cite{manchester}. The detectors and the fan-ins are kept in
place through vacuum chucks on the assembly station and their positioning is
achieved through four fiducial marks on the turn plate, shown at the centre of
Fig.~\ref{fig:assembly:b}. After the alignment, the detectors, the fan-ins and
the hybrid are mounted on the turn plate, by means of the transfer plates
(Fig.~\ref{fig:assembly:b}, top left and right), and they are then transferred
to the glue dispenser (Fig.~\ref{fig:assembly:c}).

\begin{figure}[ht]
  \centering
  \subfloat[Alignment and assembly stand.]{
    \label{fig:assembly:a}
    \epsfig{file=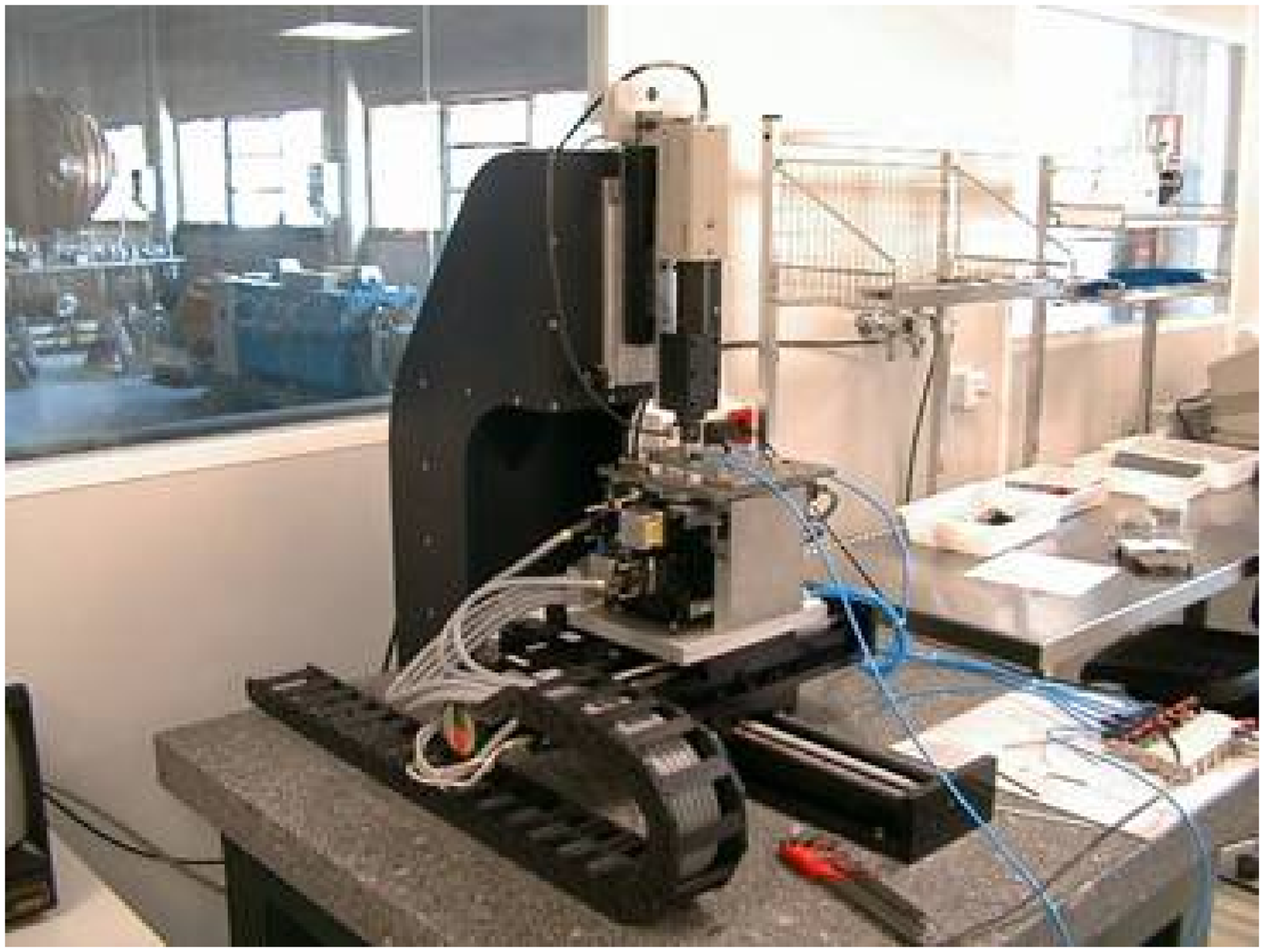,width=0.37\linewidth,clip=}
  }\hspace{0.01\linewidth}
  \subfloat[Assembly tooling.]{
    \label{fig:assembly:b}
    \epsfig{file=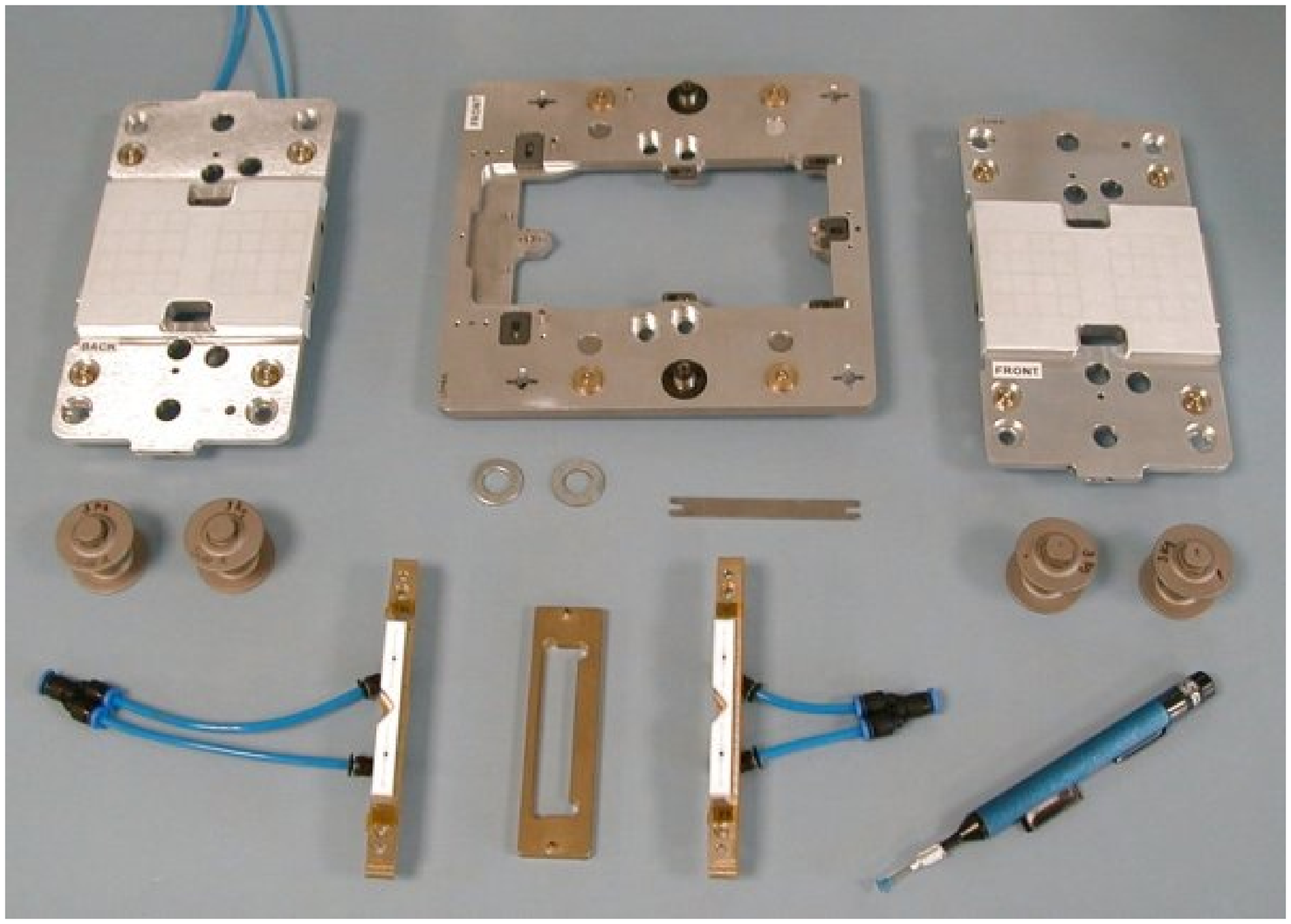,width=0.32\linewidth,clip=}
  }\hspace{0.01\linewidth}
  \subfloat[Dispenser machine.]{
    \label{fig:assembly:c}
    \epsfig{file=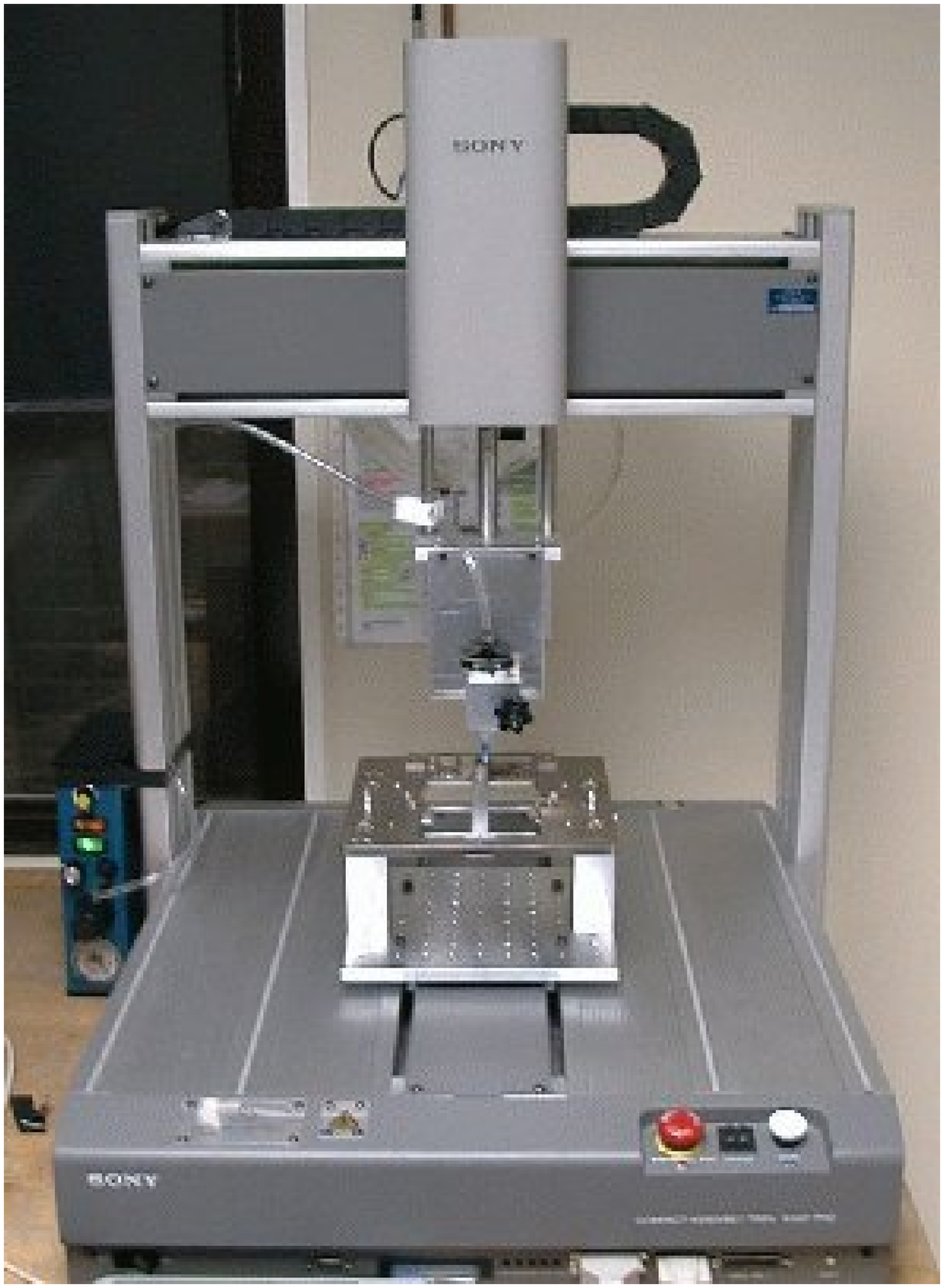,width=0.23\linewidth}
  }\caption{Alignment, assembly and gluing apparatuses for module building.} \label{fig:assembly}
\end{figure}

\item[Glue dispensing.]
The glue dispensing device, shown in Fig.~\ref{fig:assembly:c}, works on a Sony
Cast Pro system with a dispensing system attached and controlled by a PC using
the LUNA language. The thermally conductive glue applied is a two-component,
room-temperature curing epoxy\footnote{Araldite~2011, supplied by Ciba-Geigy.}
\cite{components}. The mixture makes an adhesive of adequate strength and high
thermal conductivity.

\item[Detector assembly and glue curing.]
After applying the glue, each transfer plate consecutively is inserted in and
fastened to the turn plate at the same place where it was when the aligned
detectors were picked up from the assembly station. The module components are
then left to cure in a four-pillar stand for several hours.

\item[\boldmath Detector $IV$ scan and alignment check.\unboldmath]
The measurement of the electrical properties of sensors and the visual
inspection of the components \cite{FDRQA} takes place in a probe station
comprising a microscope and a position controller installed in a dark box, as
shown in Fig.~\ref{fig:bonder:a}. The electrical properties of the silicon
sensors were examined using the home-designed program \emph{Probe\/}
\cite{probe}. It is a C++ program using the graphical user interface package
\emph{Qt}, running under Linux, which controls the measurement
instruments\footnote{Keithley~237 HV source measure unit; Keithley~236 source
measure unit; Wayne Kerr 6425 Precision Component Analyzer; Pickering
$12\times4$ switching matrix; and Keithley~2700 multimeter.} via an IEEE488
interface.

\item[Hybrid assembly with fan-ins and washers.]
The hybrid \cite{hybrid} is lowered onto its correct position on the turn plate
by inserting small tapered pins through the hybrid holes. Conductive glue is
applied underneath the HV tongues to ensure connection to the detector
back-planes. These electrical connections and the absence of short circuits is
checked before going to the next step. The fan-ins \cite{components} are added
later using the fan-ins chucks, shown in Fig.~\ref{fig:assembly:b} (bottom),
and the glue dispensing machine. The fan-ins chucks are positioned using
location pins of the proper side of the turn plate and they are tightened with
screws. The procedure is completed with the fitting of the location washers
\cite{components} on the module.

\item[\boldmath $XY$ and $Z$ survey.\unboldmath]
The setup and measuring procedure \cite{metrology} are described in detail in
Sec.~\ref{sec:metrology}. After the first metrology survey, the HV line is
glued and left to cure.

\item[Wire-bonding.]
All the production modules are wire-bonded by the fully automatic bonder
machine\footnote{K\&S~8090.} pictured in Fig.~\ref{fig:bonder:b}. The
laboratory is also equipped with a semi-automatic bonder\footnote{K\&S~1470.}
used during the pre-qualification period. A pull-tester,\footnote{Dage~3000}
shown in Fig.~\ref{fig:bonder:c}, was also used during the qualification period
for the validation of the bonding parameters.

\begin{figure}[h]
  \centering
  \subfloat[Probe station.]{
    \label{fig:bonder:a}
    \epsfig{file=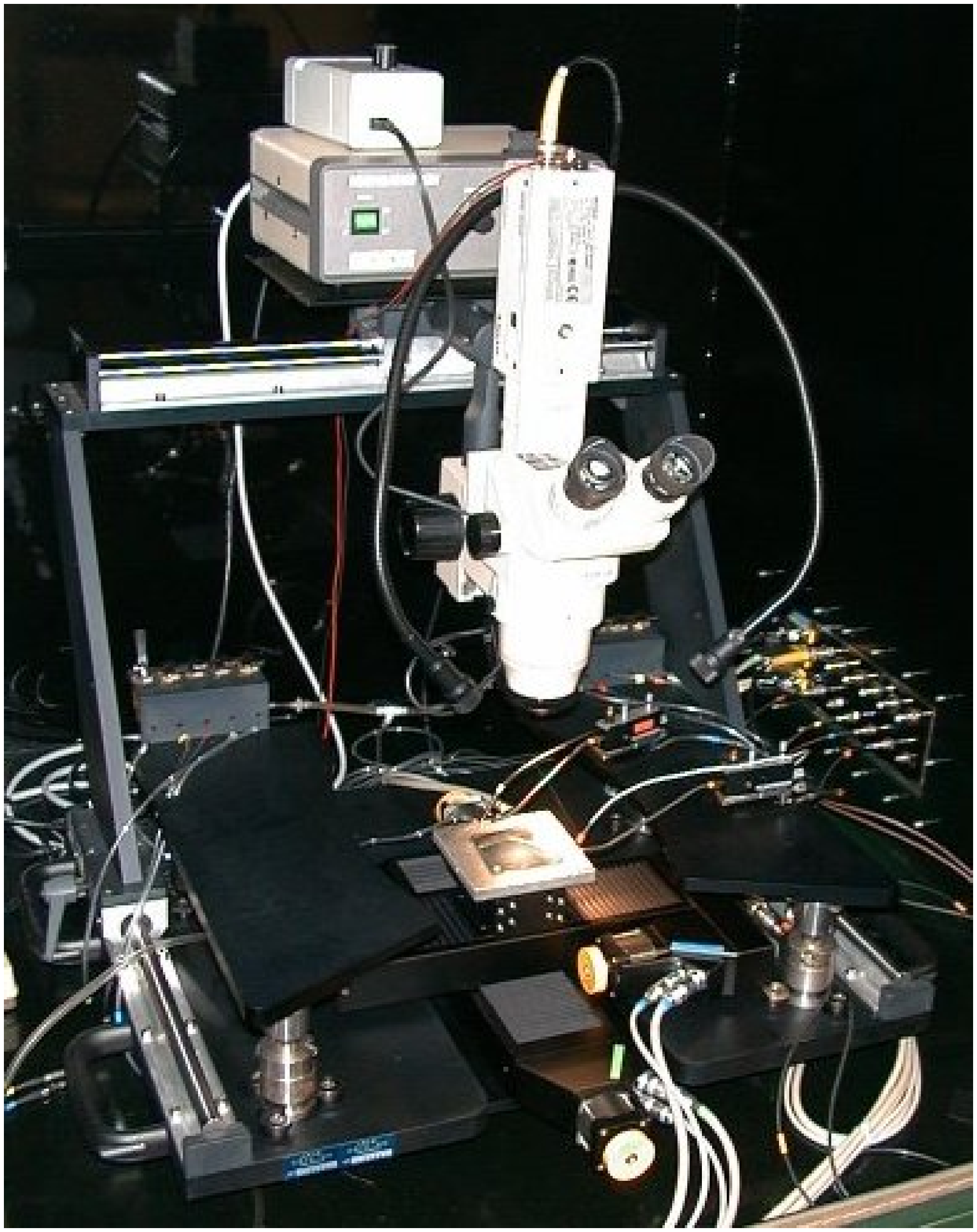,width=0.3\linewidth}
  }\hspace{0.01\linewidth}
  \subfloat[Automatic bonding device.]{
    \label{fig:bonder:b}
    \epsfig{file=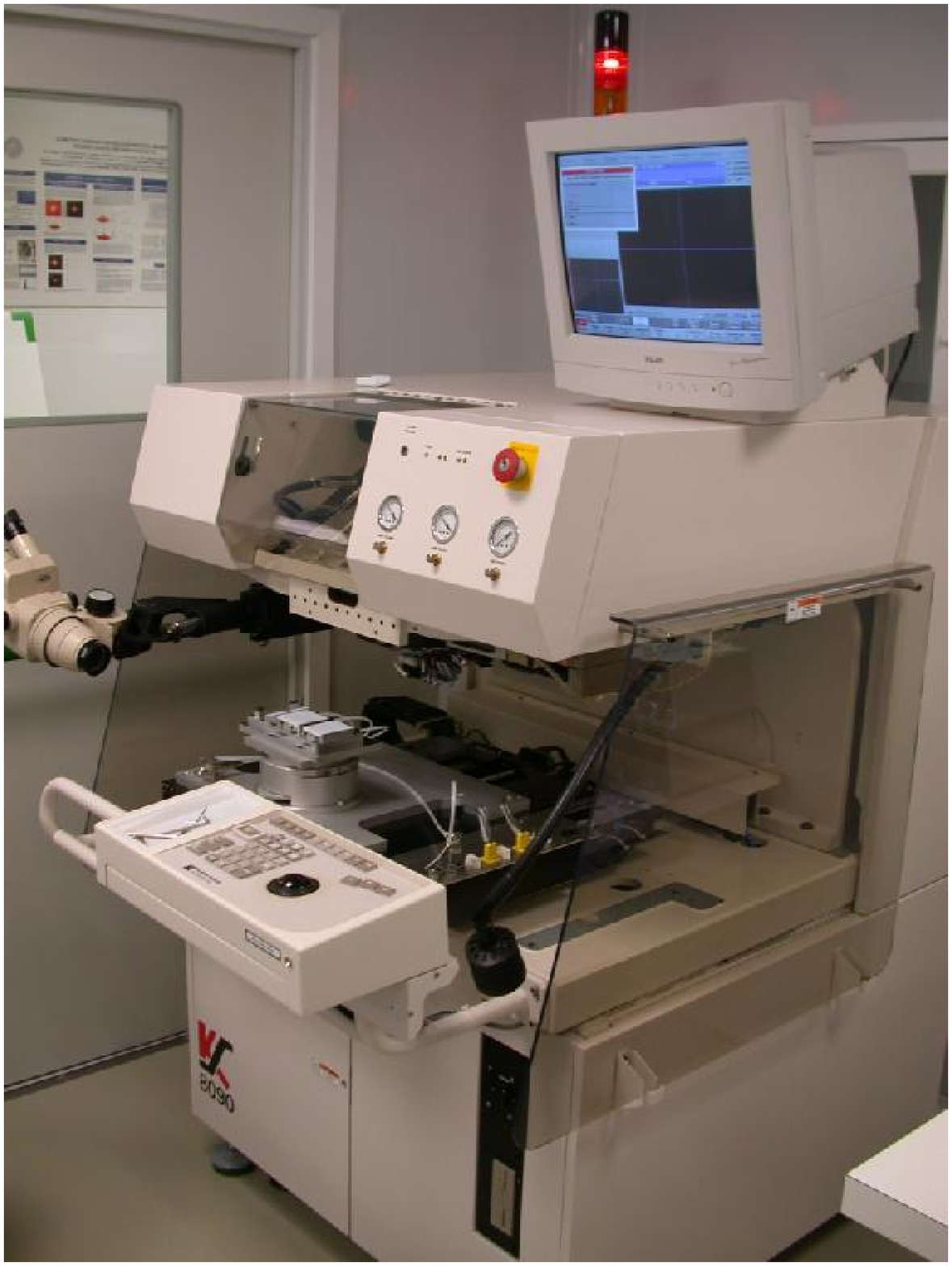,width=0.32\linewidth}
  }\hspace{0.01\linewidth}
  \subfloat[Pull-testing machine.]{
    \label{fig:bonder:c}
    \epsfig{file=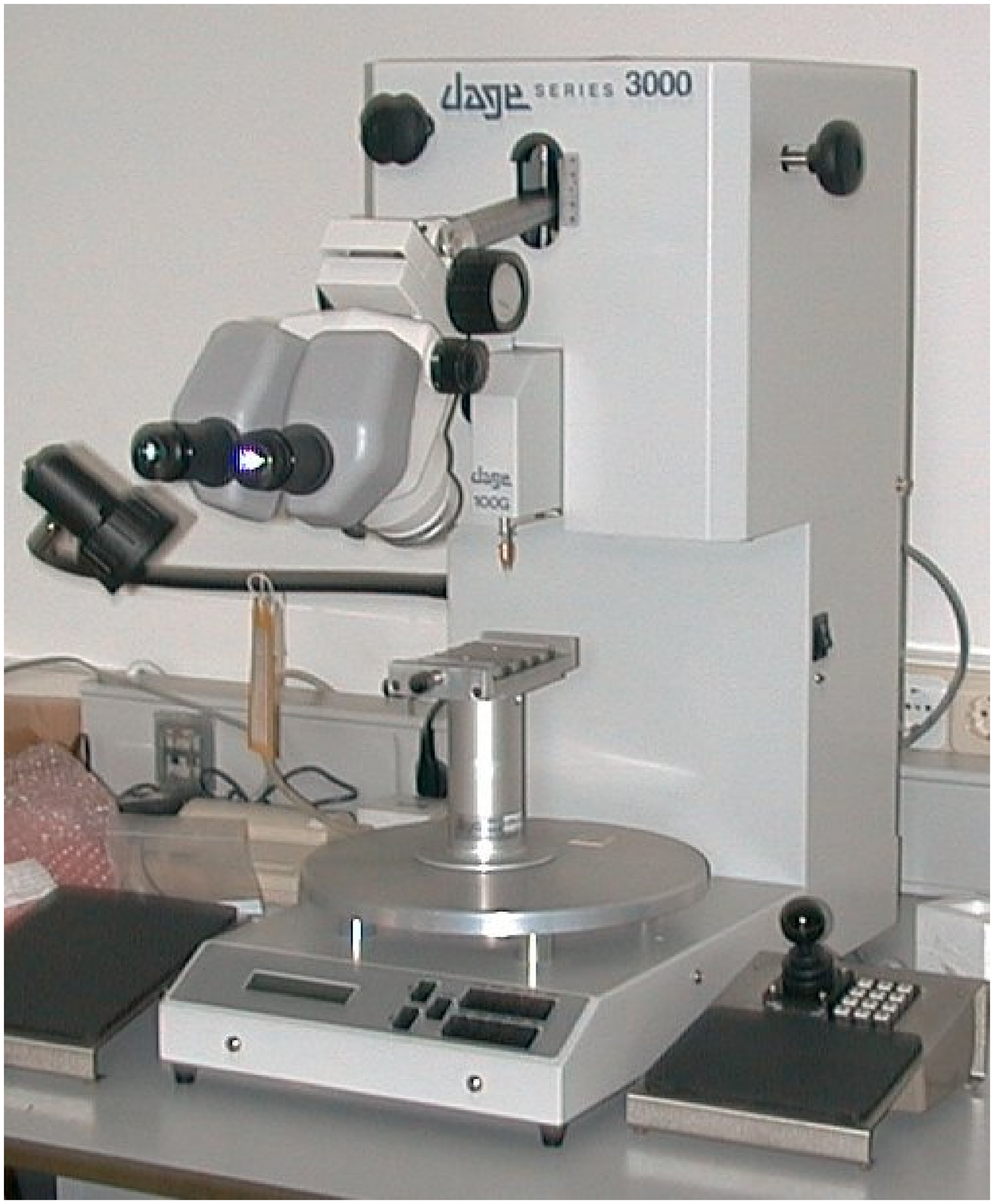,width=0.3\linewidth}
  }\caption{Devices for module testing and wire-bonding.}\label{fig:bonder}
\end{figure}

\item[Thermal cycling.]
When the assembly is completed, the modules undergo a thermal cycling in order
to verify that temperature variations do not compromise the geometrical
properties of the module \cite{mech_specs}. This task is performed in a climate
chamber,\footnote{Dycometal CM~40/125A.} displayed in Fig.~\ref{fig:thermo:a},
which operates at a temperature ranging from $\rm-30~^{\circ}C$ to
$\rm+50~^{\circ}C$. An aluminium rack positioned inside the chamber, shown in
Fig.~\ref{fig:thermo:b}, allows simultaneous thermal cycling of up to six
modules. The thermistors are readout by a multimeter\footnote{Keithley~2700.}
and the system is controlled, monitored and read out by a PC. The thermistors
temperature was recorded every 10~s in a data file and plotted on-line (cf.\
Fig.~\ref{fig:thermo:c}). During this process the thermistors temperature
varies from $\rm-30~^{\circ}C$ to $\rm+35~^{\circ}C$ for a total of ten cycles
lasting about 3.5~h each.

\begin{figure}[ht]
  \centering
  \subfloat[Thermocycling setup.]{
    \label{fig:thermo:a}
    \epsfig{file=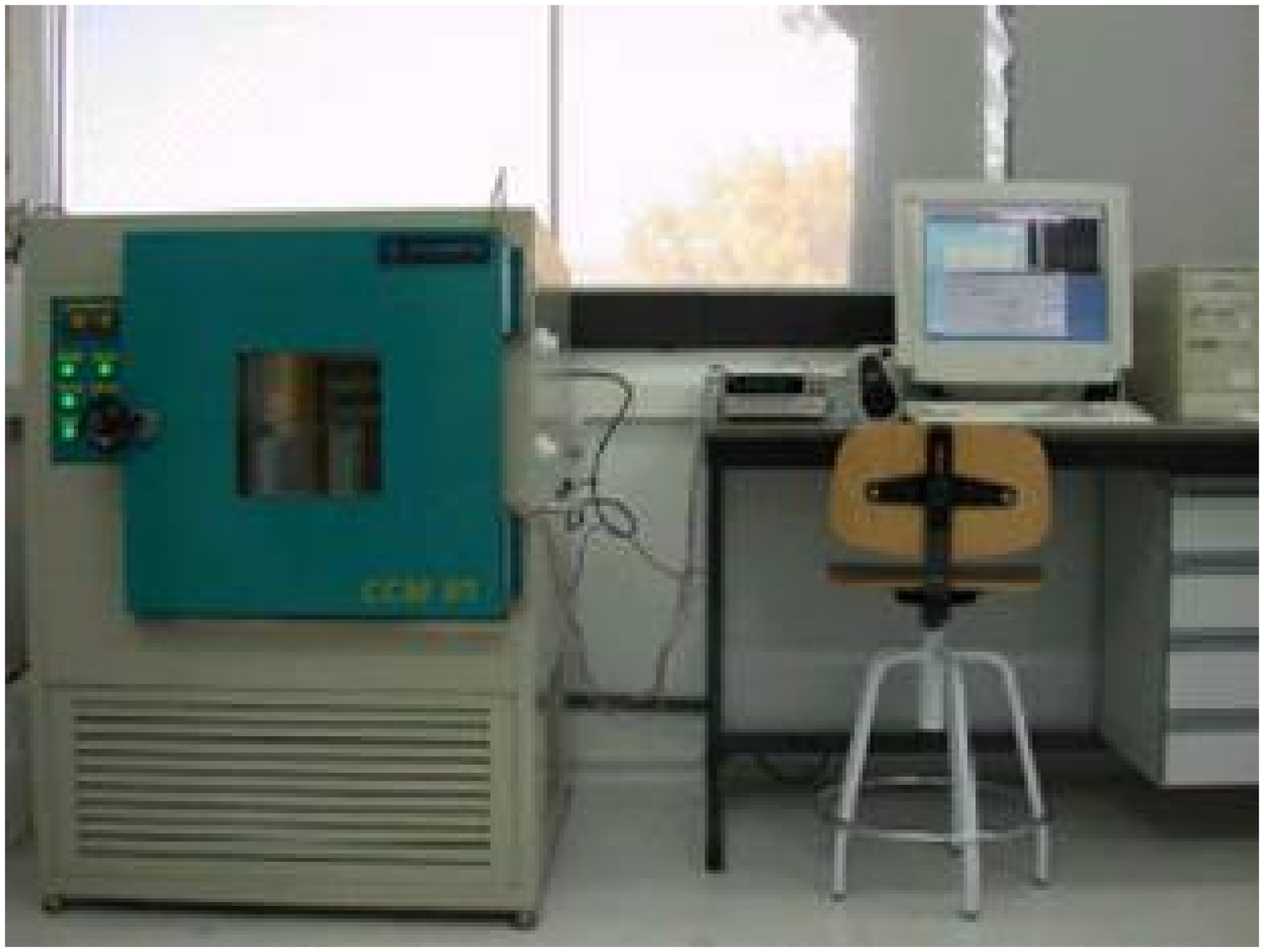,width=0.28\linewidth,clip=}
  }\hspace{0.005\linewidth}
  \subfloat[Interior of climate chamber.]{
    \label{fig:thermo:b}
    \epsfig{file=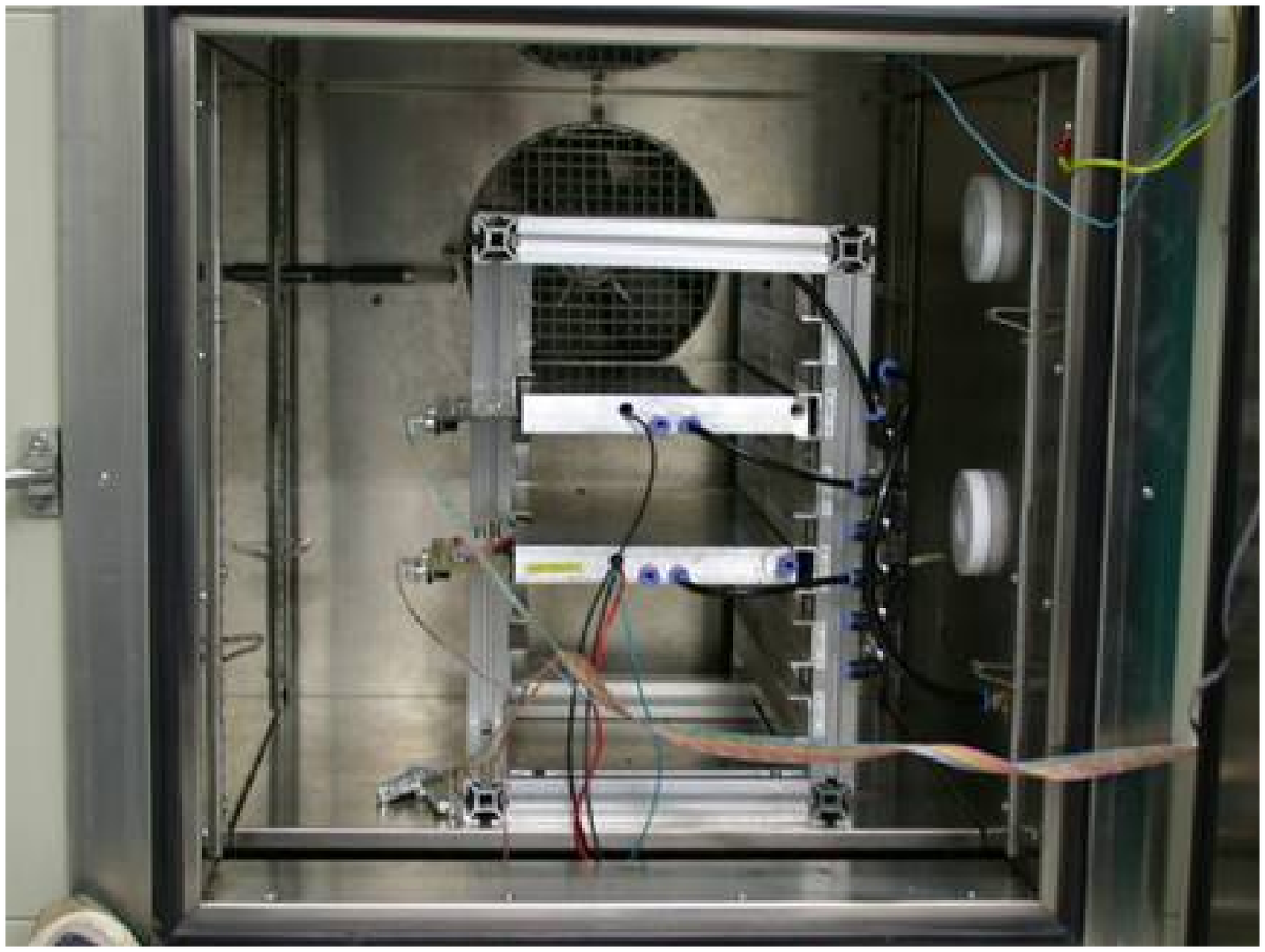,width=0.28\linewidth}
  }\hspace{0.005\linewidth}
  \subfloat[Thermal cycle.]{
    \label{fig:thermo:c}
    \epsfig{file=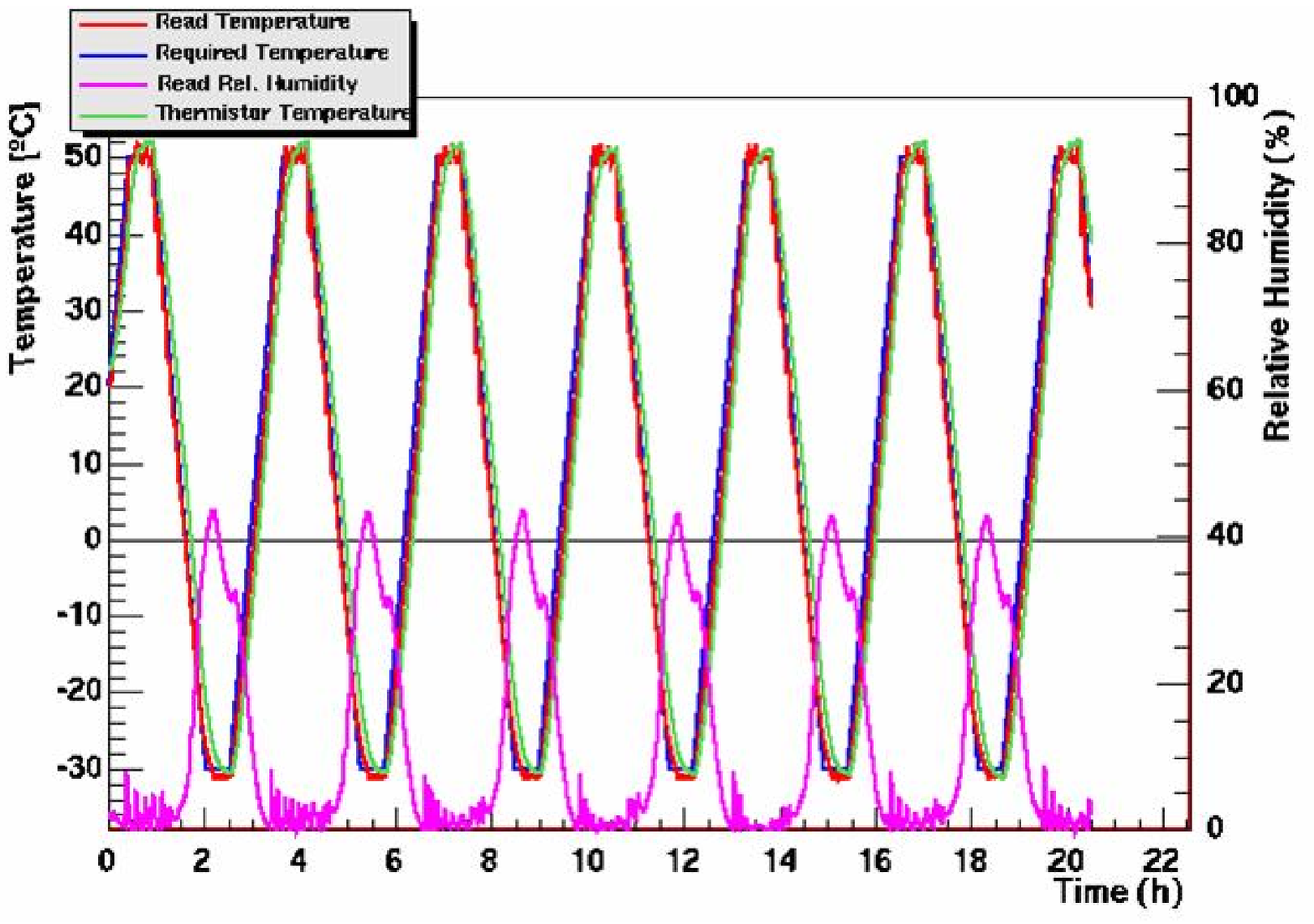,width=0.36\linewidth}
  }
\caption{The thermal cycling setup and a typical monitoring plot. The oven and
thermistor temperatures are plotted, together with the relative humidity. A
complete process covers ten cycles.}\label{fig:thermo}
\end{figure}

\end{asparadesc}

The quality assurance plan for the forward modules foresees the following QC
tests \cite{FDRQA} for the completed (glued and bonded) module:
\begin{compactitem}
  \item final $XY$ and $Z$ metrology \cite{metrology};
  \item long-term leakage current and electrical stability \cite{elec_procs};
  \item electrical characterisation \cite{elec_procs};
  \item $IV$ curve of the completed module \cite{elec_procs};
  \item visual inspection.
\end{compactitem}

In the following two sections, we concentrate on the procedures and outcome of
the most important quality control issues: the dimensions in
Sec.~\ref{sec:metrology} and the electrical behaviour in
Sec.~\ref{sec:electrical} of the produced modules.

\section{Metrology}\label{sec:metrology}

The precision needed during construction is determined by physics requirements.
For the detector to be hermetic in tracks with $p_{\rm T}>1~{GeV}$, a tolerance
of \mbox{$\sim100~\mu m$} in the $r\phi$ direction and \mbox{$\sim500~\mu m$}
in radius is specified \cite{mech_specs}. Taking this into consideration, the
effective module dimensions specifications are derived \cite{metrology}.

\subsection{Setup and procedures}\label{sub:metro_specs}

The metrology process is divided into two main tasks (see Ref.~\cite{metrology}
for a detailed description); the $XY$ metrology,\footnote{The $x$-axis is
defined along the strips and the $y$-axis is perpendicular to them and on the
sensor plane.} in which relative in-plane alignment between silicon sensors and
the mounting hole and slot\footnote{As hole is denoted the mounting point
between the hybrid and the sensors, whereas the slot is located in the module
endpoint.} are measured, and the $Z$-profile of the detectors relative to the
mounting surface. The position of sensors in the $xy$-plane is defined by a
number of fiducial marks printed on its surface. The hole and mounting slot
have a sharp and well-defined edge which is used for determining its position.

The $XY$ metrology parameters, which refer to the relative positions of the
four detectors in the $xy$-plane with respect to each other, are the following:
\begin{compactitem}
  \item $midxf$, $midyf$, which represent the front-to-back detector pair
alignment in $x$ and $y$;
  \item $sepf$ and $sepb$, which represent the separation between the
detectors of each side;
  \item $stereo$ angle, which denotes the deviation from the nominal stereo angle
($\rm40~mrad$) between the two detector pairs;
  \item $a1$, $a2$, $a3$, and $a4$, which are the four individual detector angles
relative to their nominal position; and
  \item $mhx$, $mhy$, $msx$, $msy$, are the hole and slot positions in $x$ and
$y$, respectively.
\end{compactitem}
The specified nominal values and the corresponding tolerances for these
parameters are listed in Table~\ref{tab:XYspecs}.

\begin{table}[ht]
\begin{center}
\begin{tabular}{ l c c c }                 \hline
  Parameter [unit]\T\B & Nominal value (middle) & Nominal value (outer) & Tolerance \\
  \hline
  $mhx$ [mm]\T\B      & \phm71.708    & -78.136       & 0.020 \\
  $mhy$ [mm]          & \phm\pho0.000 & \phm\pho0.000 & 0.020 \\
  $msx$ [mm]          & -66.672       & \phm62.244    & 0.100 \\
  $msy$ [mm]          & \phm\pho0.000 & \phm\pho0.000 & 0.020 \\
  $midxf$ [mm]        & \phm\pho0.000 & \phm\pho0.000 & 0.010 \\
  $midyf$ [mm]        & \phm\pho0.053 & \pho-0.040    & 0.005 \\
  $sepf$, $sepb$ [mm] & \phm59.900    & \phm61.668    & 0.010 \\
  $a1$--$a4$ [mrad]   & \phm\pho0.000 & \phm\pho0.000 & 0.130 \\
  $stereo$ [mrad]     & -20.000       & -20.000       & 0.130 \\ \hline
\end{tabular}
\caption{$XY$ parameters nominal values and tolerances.} \label{tab:XYspecs}
\end{center}
\end{table}

\begin{figure}[ht]
  \centering
  \epsfig{file=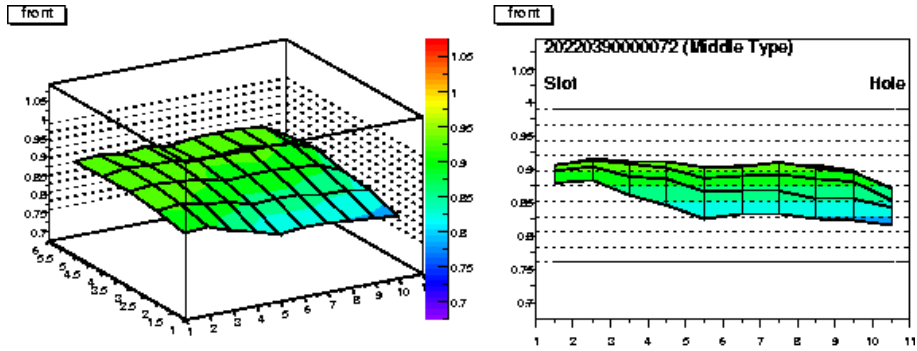,width=0.8\linewidth,bbllx=14,bblly=140,bburx=332,bbury=261,clip=}
\caption{Typical $Z$ profile of the front side of a middle module. The
back-side profile is obtained likewise.} \label{fig:M72_Z_afterTC}
\end{figure}

The $Z$ metrology,\footnote{The $z$-axis is defined vertical to the module.} on
the other hand, is performed by measuring the $Z$ position of a grid of
25~points distributed over the surface of each of the four detectors, as shown
in Fig.~\ref{fig:M72_Z_afterTC}. From these points, the average ($zave$), the
maximum ($zmax$), the minimum ($zmin$) and the RMS ($zrms$) values are
calculated for the front and back side of the module. These parameters are then
compared with the specifications, which are listed in Table~\ref{tab:Zspecs}
\cite{metrology}.

\begin{table}[ht]
\begin{center}
\begin{tabular}{ l c c } \hline
  Module side\T\B & Nominal value (mm) & Tolerance (mm)\\
  \hline
  Front\T\B       & \phm0.875          & 0.115 \\
  Back            & -0.375             & 0.115 \\ \hline
\end{tabular}
\caption{Nominal values and tolerances for the $Z$ measurements.}
\label{tab:Zspecs}
\end{center}
\end{table}

\begin{figure}[ht]
\centering
\begin{minipage}[c]{0.39\linewidth}
  \centering\epsfig{file=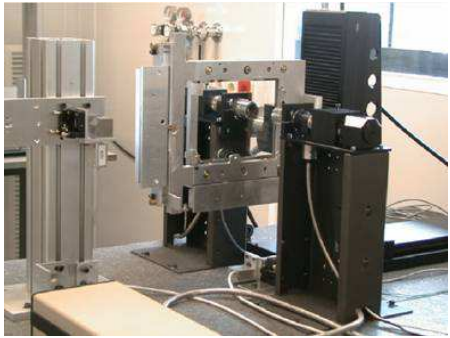,width=\linewidth}
\end{minipage} \hspace{0.05\linewidth} 
\begin{minipage}[c]{0.35\linewidth}
  \centering
\caption{The metrology setup: the two cameras, the mechanical stages and the
module-mounting frame are visible.}\label{fig:metrology}
\end{minipage}
\end{figure}

A custom-made\footnote{AIDO-IFIC.} laser interferometry metrology system, shown
in Fig.~\ref{fig:metrology} was employed for the mechanical survey of the
module, providing an accuracy of $\rm1~\mu m$. It is equipped with two cameras
to achieve a better accuracy in the determination of the $xy$ parameters. Thus
it avoids errors due to intermediate manipulation steps when turning the module
upside down, as is needed with single camera systems. During the survey, the
module is mounted in a frame allowing the measurement of the $xy$ and $z$
positions of the detectors of both sides with respect to the same reference
points.

\subsection{Measurements}\label{sub:metro_res}

\begin{figure}[ht]
\centering
\begin{minipage}[c]{0.46\linewidth}
  \centering\epsfig{file=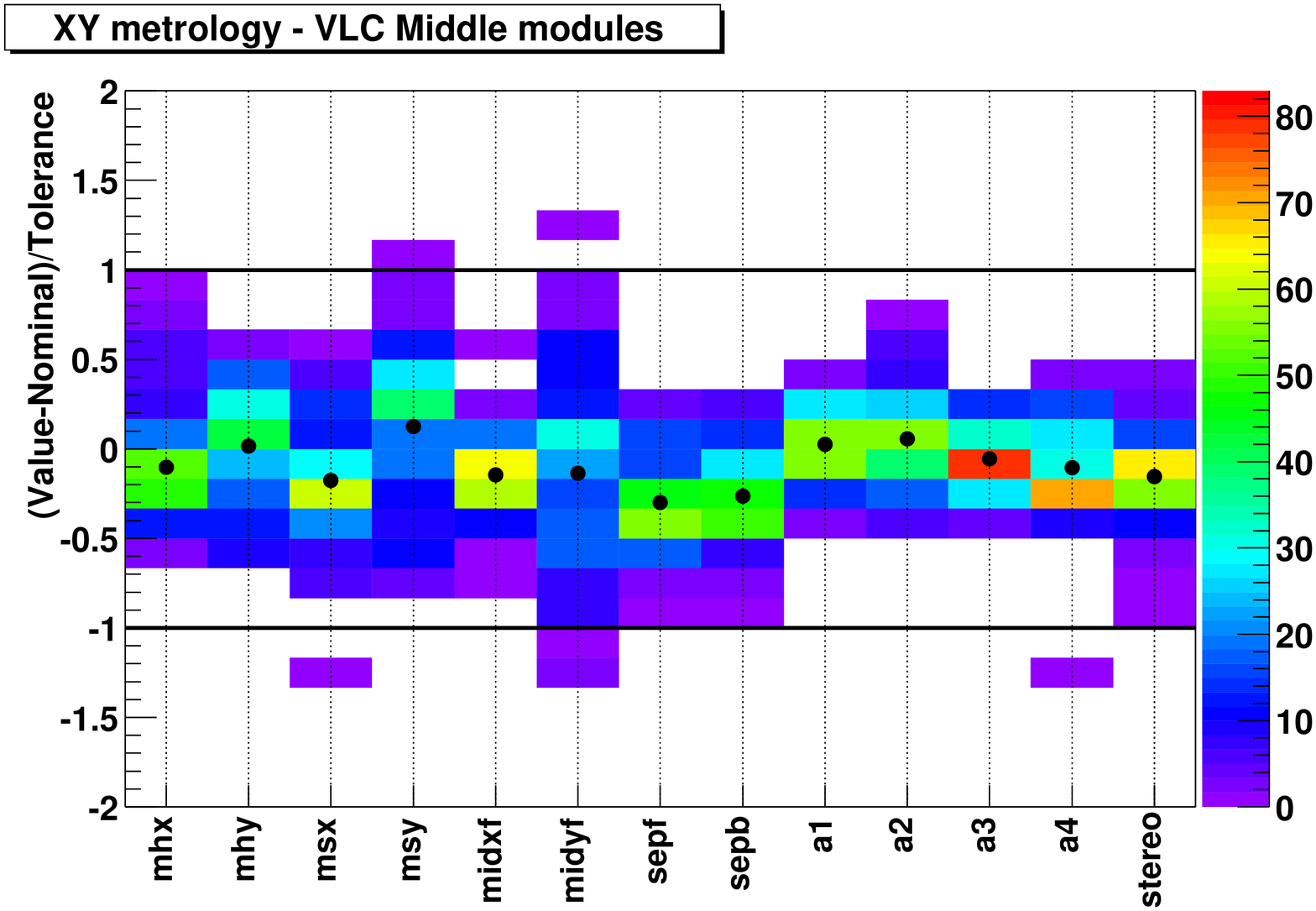,width=\linewidth}
\end{minipage} \hspace{0.03\linewidth}
\begin{minipage}[c]{0.46\linewidth}
  \centering\epsfig{file=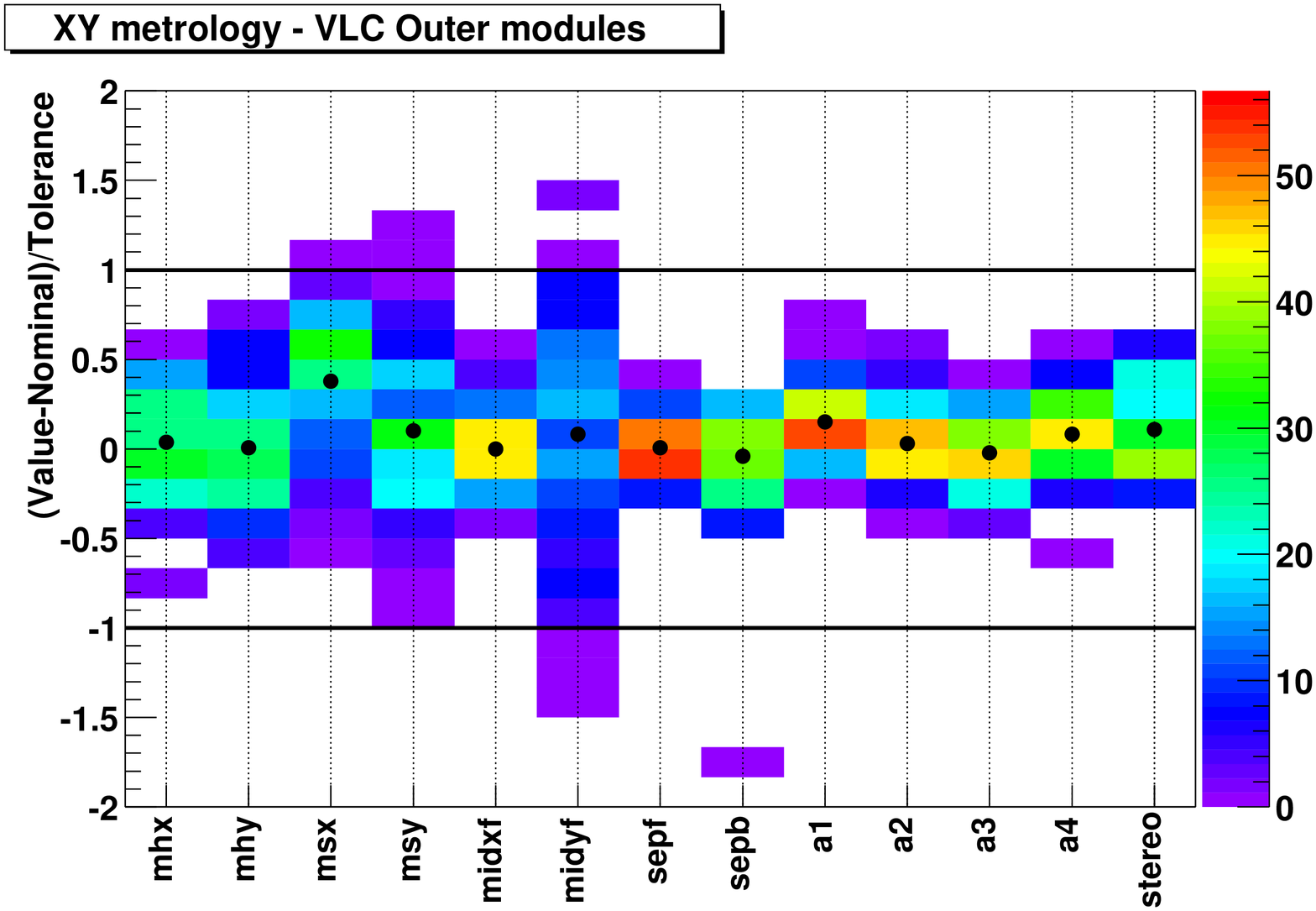,width=\linewidth}
\end{minipage}
  \caption{$XY$-parameters distributions for middle (left) and outer (right) modules,
  normalised as (\emph{value -- nominal}) $/$ \emph{tolerance}.}\label{fig:xy-all-dist}
\end{figure}

As mentioned before, the module dimensions were measured twice: just after
gluing the components together and before wire-bonding, and once more after the
thermal cycling. The overall final-measurement results for all 13~$XY$
parameters are shown in Fig.~\ref{fig:xy-all-dist}. The majority of the modules
are well within the specifications, apart from some cases where parameter
$midyf$ exceeds the allowed limits. These cases stand for the majority of the
modules characterised as \emph{Pass.\/}

\begin{figure}[ht]
\centering
\begin{minipage}[c]{0.46\linewidth}
  \centering\epsfig{file=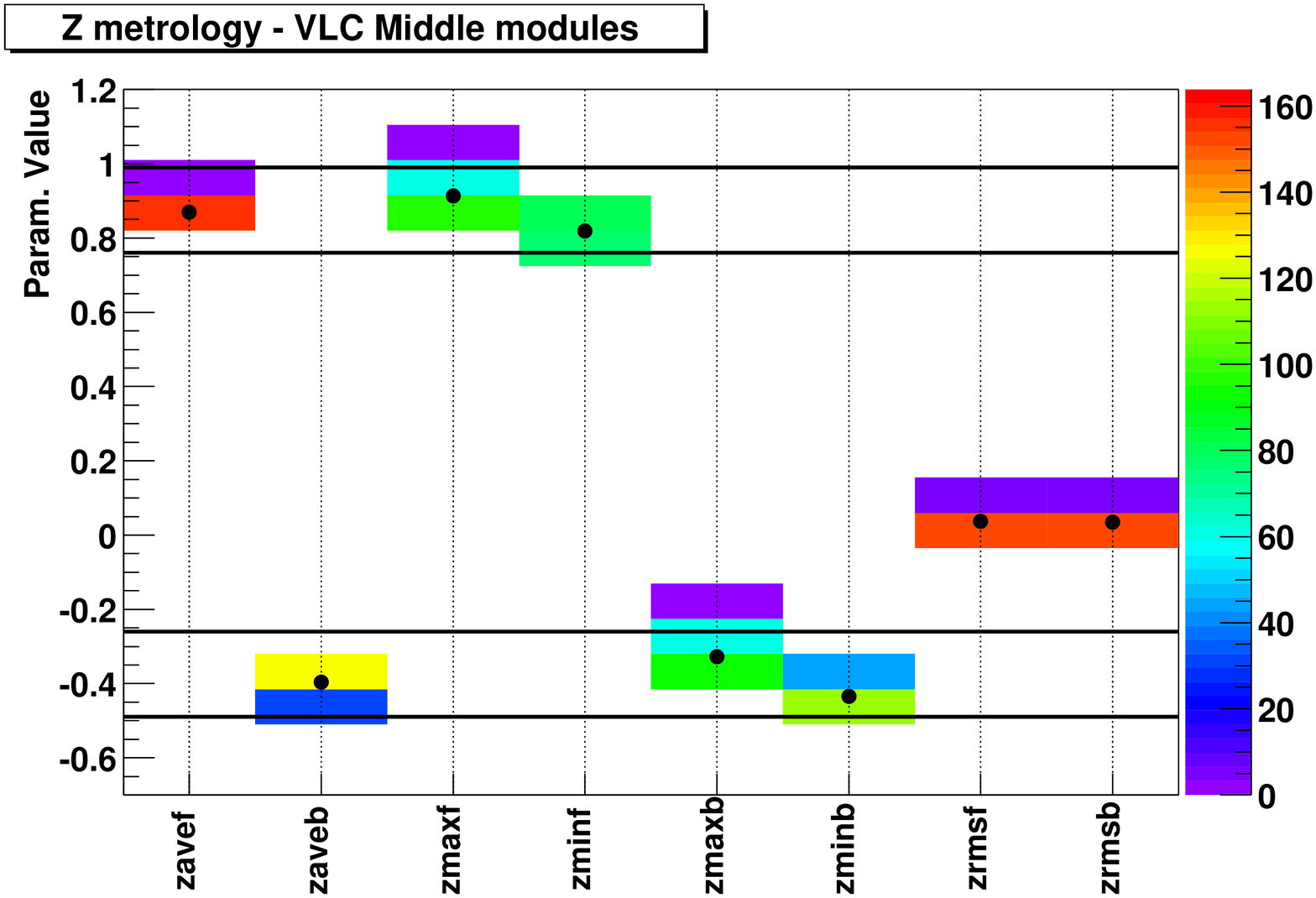,width=\linewidth}
\end{minipage} \hspace{0.03\linewidth}
\begin{minipage}[c]{0.46\linewidth}
  \centering\epsfig{file=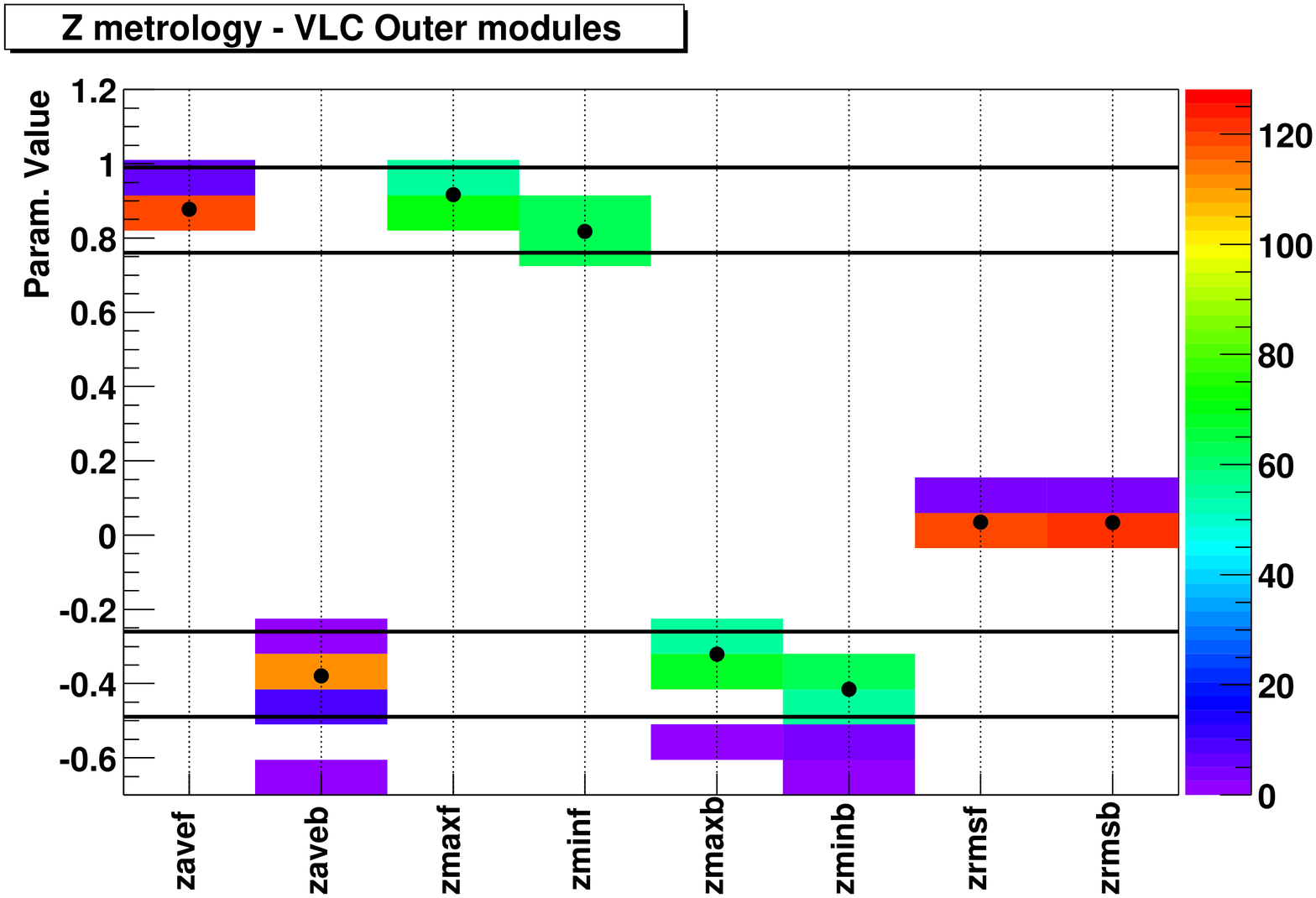,width=\linewidth}
\end{minipage}
  \caption{$Z$-parameters distributions for middle (left) and outer (right) modules,
  expressed in millimetres.}\label{fig:z-all-dist}
\end{figure}

The thermal cycling did not affect the $XY$ metrology parameters, whereas, in
some cases, it altered the $Z$ profile. Parameters $zave$, $zmin$ and $zmax$
for both sides ---in particular $zminb$--- decreased after the thermal
treatment in several modules, namely they appeared slightly bent in the centre.
However, this effect, did not have any impact whatsoever on the production
yield. The distributions of the $Z$ parameters are presented in
Fig.~\ref{fig:z-all-dist}.

\section{Electrical performance}\label{sec:electrical}

The readout of the SCT modules is based on 12~ABCD3TA \cite{abcd} ASICs mounted
on a copper/kapton hybrid \cite{hybrid}. The ABCD3TA chip features a
128-channel analog front end consisting of amplifiers and comparators and a
digital readout circuit operating at a frequency of 40.08~MHz. It utilises the
binary scheme, where the signals from the silicon detector are amplified,
compared to a threshold and only the result of the comparison enters the input
register and a 132-cell deep pipeline, awaiting a level-1 trigger accept
signal. It implements a redundancy mechanism that redirects the output and the
control signals, so that a failing chip can be bypassed. To reduce the
channel-to-channel threshold variation, in particular after irradiation, the
ABCD3TA features an individual threshold correction in each channel with a
4-bit digital-to-analog converter \emph{(TrimDAC)\/} with four selectable
ranges. The clock and command signals as well as the data in binary form are
normally transferred from and to the off-detector electronics through optical
links \cite{optical}, however during the production electrical tests, the ASICs
were read out electrically using a pseudo-optical scheme. Two streams of data
are read out, one for each module side (768~channels).

In this section the electrical specifications and the respective quality
control procedures are outlined. Among the various functionality tests of the
readout part of the module, only the more crucial ones such as the gain and the
noise measurements are elaborated. Some exceptional cases are also discussed in
detail. A complete description of the electronics test procedure is available
in Ref.~\cite{elec_procs}.

\subsection{Specifications and setup} \label{sub:elec_specs}

The LHC operating conditions demand challenging electrical performance
specifications for the SCT modules and the limitations mainly concern the
accepted noise occupancy level, the tracking efficiency, the timing and the
power consumption. The Equivalent Noise Charge (ENC) is specified
\cite{elec_specs} to be less than 1500~electrons before irradiation. The noise
hit rate is required to be $<5\times10^{-4}$ per strip at the ATLAS SCT
operating threshold of 1~fC. At this threshold an efficiency of 99.5\% is
expected before irradiation, however a threshold higher than 1~fC might be
needed after irradiation to meet the efficiency and noise occupancy
specifications \cite{endcap}. As far as the tracking performance is concerned,
a starting requirement is a low number of dead readout channels, specified to
be less than~16 for each module to assure at least 99\% of working channels.
Furthermore no more than eight consecutive faulty channels are accepted in a
module.

\begin{figure}[ht]
  \centering\epsfig{file=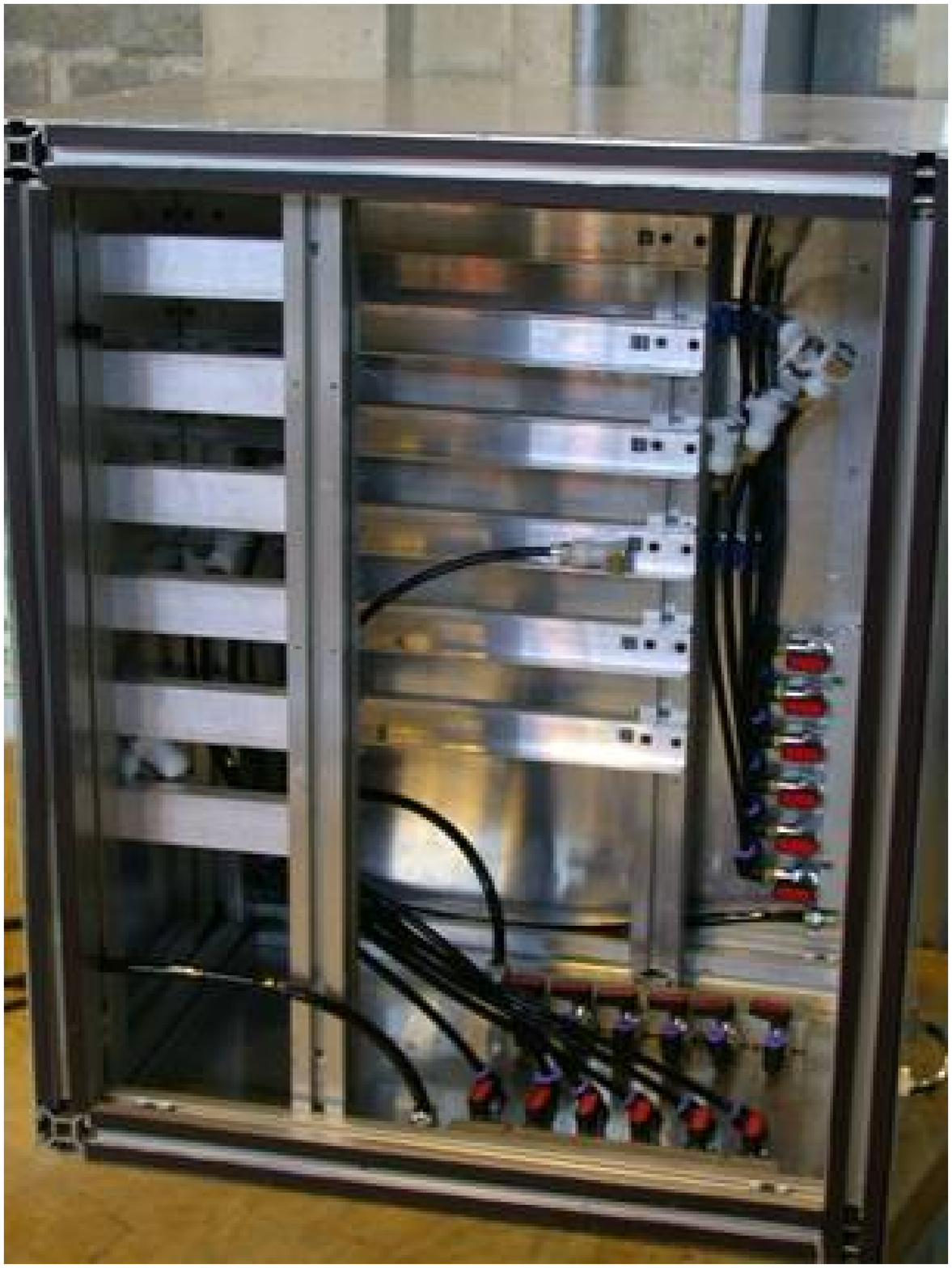,width=0.26\linewidth}
  \hspace{0.2\linewidth} 
  \epsfig{file=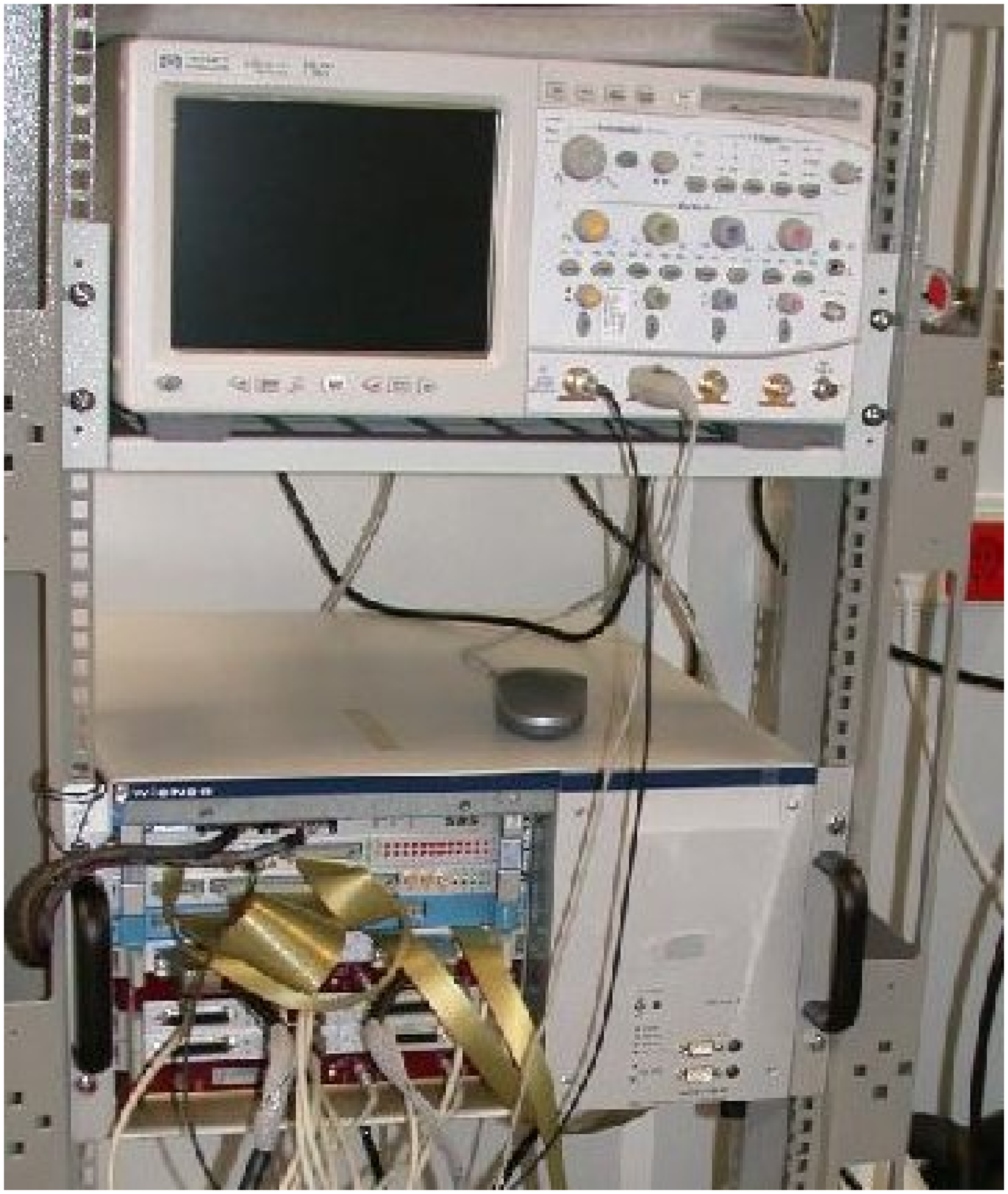,width=0.295\linewidth} \caption{Electrical tests setup: the light-proof box (left) and the
readout boards installed in the VME crate (right).} \label{fig:elec_setup}
\end{figure}

A standard DAQ system has been developed within the ATLAS SCT collaboration
aiming at verifying the hybrid and detector functionality after the module
assembly and at demonstrating the module performance with respect to the
required electrical specifications. During the electrical tests the modules are
mounted in a light-tight aluminum box which supports the modules at the two
washers. The test box includes a cooling channel connected to a liquid coolant
system.\footnote{Huber chiller with antifreeze as a coolant.} The operating
temperature is monitored by a thermistor mounted on the hybrid. The box also
provides a connector for dry air circulation. Subsequently, the module test box
is placed inside the custom-made, light-proof box, shown in
Fig.~\ref{fig:elec_setup} (left), and it is electrically connected to the
readout system and VME crate. Up to six modules can be tested simultaneously
with this configuration. The grounding and shielding scheme of the setup is of
crucial importance, therefore a careful optimisation is necessary.

In all the measurements performed, the ASICs are powered with three standard
SCT low-voltage power supplies, SCTLV, and read out electrically via a system,
comprising one SLOG, one MUSTARD and one AERO board, installed in two VME
crates as pictured in Fig.~\ref{fig:elec_setup} (right). Two SCTHV high voltage
units provided detector bias to the modules. A second complete setup including
the respective readout system was also built, to serve as a backup in the event
of failure of the first one. A detailed description of the SCT readout system
is given in Refs.~\cite{elec_procs,endcap}.

The software package SCTDAQ \cite{sctdaq} has been implemented for testing both
the bare hybrids and the modules using VME units. It consists of a C++
dynamically linked library and a set of ROOT \cite{root} macros which analyse
the raw data obtained in each test and stores the results in a database
\cite{DB}.

\subsection{Testing sequence}\label{sub:elec_tests}

To determine the front-end parameters of the modules, an internal calibration
circuit that simulates an input charge in the range $\rm0.5-10~fC$ is
implemented in the ASICs. By injecting various known charges and performing
threshold scans, the analogue properties of each channel can be determined,
such as the gain, the offset and the noise. A complementary error function is
fitted to each threshold scan to determine the point of 50\% efficiency
($vt_{50}$) and the output noise for each channel. A multi-parameter fit to a
set of $vt_{50}$ points is used to obtain the response curve from which gain
and offset for each channel are derived. The input noise is thus calculated by
dividing the output noise measured at $\rm2~fC$ over the calculated gain. The
testing sequence followed in Valencia is the following:

\begin{compactdesc}

  \item[Long-term test:]
This is a burn-in test, during which the ASICs are powered, clocked and
triggered for 24~hours while the module bias voltage is kept at 150~V and its
thermistor temperature is \mbox{$\rm\sim\!10~^{\circ}C$}. The bias voltage,
chip currents, hybrid temperature, the leakage current and the noise occupancy
are recorded every 15~min. Moreover, every two hours a so-called
\emph{confirmation}\footnote{It includes digital tests, a strobe delay setting
and a three-point gain.} test is performed to verify the correct functionality
of the module.

  \item[Characterisation:]
A full electrical characterisation of the module is carried out at a hybrid
operating temperature of $\rm10\pm5~^{\circ}C$ and a bias voltage of 150~V,
consisting of the following tasks:
\begin{asparaenum}[a)]
    \item \emph{Digital tests:\/}
checks of the redundancy links, the chip by-pass functionality and the 128-cell
pipeline circuit, executed in order to identify chip or hybrid damage.
    \item \emph{Strobe delay:\/}
an optimisation of the delay between the calibration signal and the clock on a
chip-to-chip basis.
    \item \emph{Three-point gain:\/}
a first evaluation of the gain for three different values of injection charge.
    \item \emph{Trimming:\/}
adjustment of the TrimDAC for each channel to allow for an improved matching of
the comparators thresholds.
    \item \emph{Response Curve:\/}
final measurement of the gain for ten values of injected charge ranging from
$\rm0.5~fC$ to $\rm8~fC$.
    \item \emph{Noise occupancy:\/}
a threshold scan without any charge injection, performed to yield a direct
measurement of the noise occupancy at the equivalent input charge of 1~fC,
which is the operating threshold for the SCT module. The adjusted discriminator
offset is applied to ensure a uniform measurement across the channels.
    \item \emph{Time walk:\/}
a strobe-delay scan is performed to assess the sensitivity of the pulse timing
to the injected charge.
\end{asparaenum}

  \item[Final \boldmath $IV$ \unboldmath scan:]
A final measurement of the detector leakage current as a function of the bias
voltage ($IV$ curve) is performed at \mbox{$\rm\sim\!20~^{\circ}C$} to assure
that the current drawn by the whole module is low enough for the safe operation
of the detector. The current values at 150~V and 350~V are recorded and
compared with those of previous $IV$ curve measurements before and after the
module sub-assembly.

\end{compactdesc}

The characterisation sequence is also applied as a reception test of the
unassembled hybrids.

\subsection{Electrical results}\label{sub:elec_results}

\begin{figure}[ht]
\begin{minipage}[c]{0.48\linewidth}
  \centering\epsfig{file=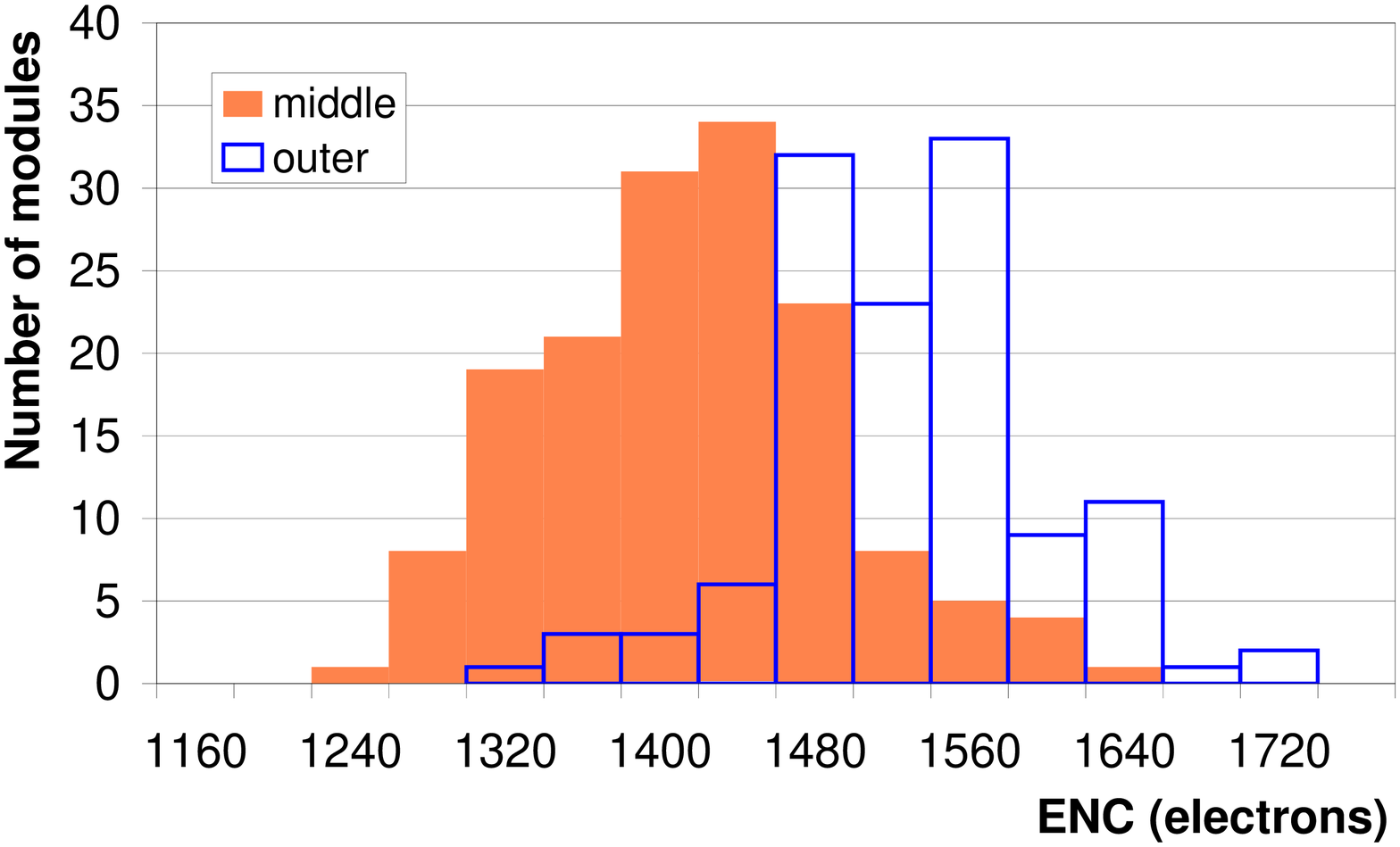,width=\linewidth,clip=}
\caption{Average equivalent noise charge distributions for middle and outer
modules.} \label{fig:enc}
\end{minipage}\hfill
\begin{minipage}[c]{0.48\linewidth}
  \centering\epsfig{file=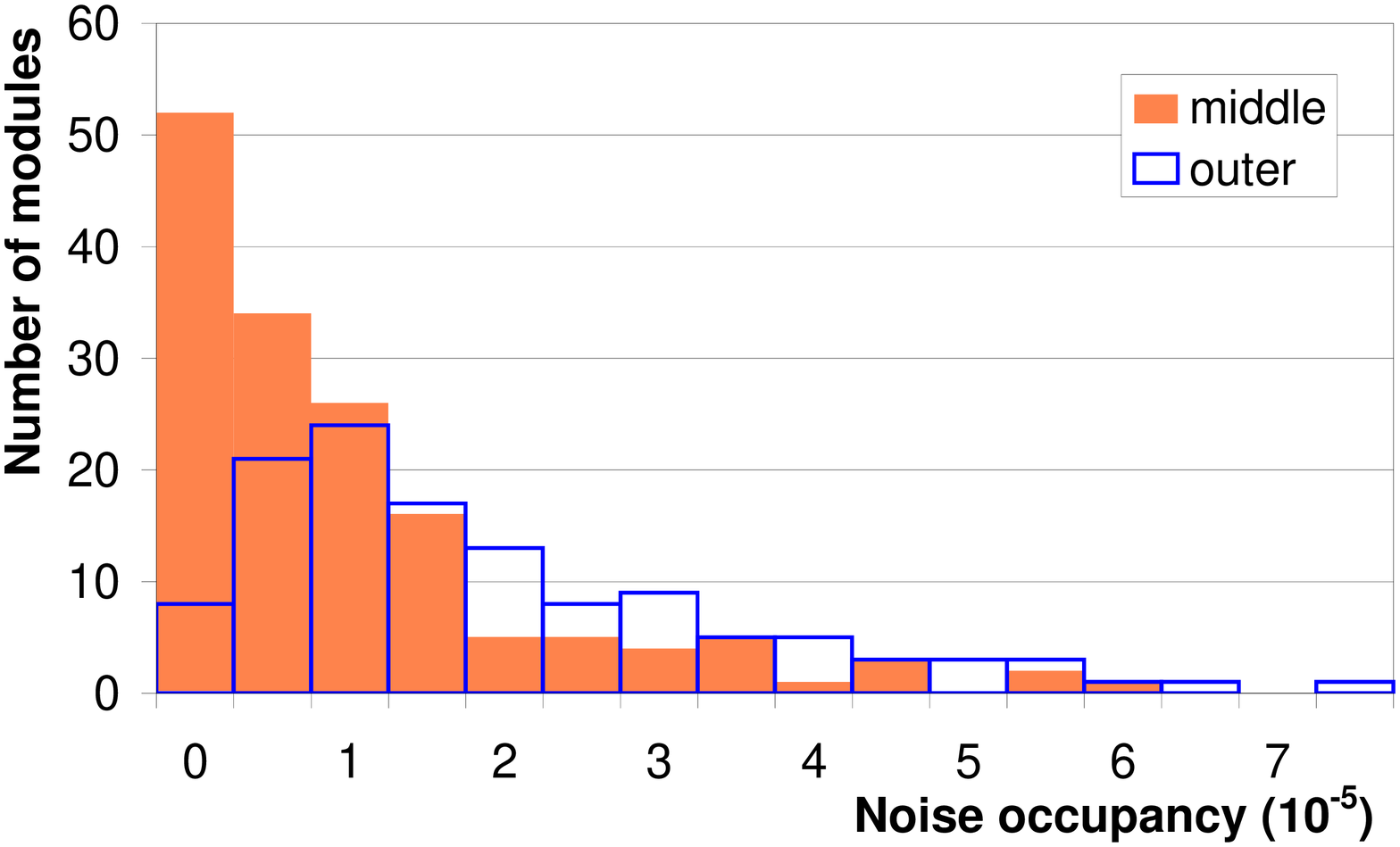,width=\linewidth,clip=}
\caption{Average noise occupancy distributions at 1~fC for middle and outer
modules.} \label{fig:NO}
\end{minipage}
\end{figure}

Collective results, as far as noise is concerned, are presented in
Figs.~\ref{fig:enc} and~\ref{fig:NO}, for middle and outer modules separately.
Middle modules, being shorter than the outer ones,\footnote{In middle modules
the total strip length is 117.7~mm, whereas in outer modules it is 121.2~mm.}
exhibit a lower noise level. The noise occupancy distributions lie well below
the specification of $5\times10^{-4}$. It is worth noting that the acquired
noise measurements largely depend on the specific setup optimisation level
(e.g.\ grounding, shielding), therefore these values represent rather an upper
limit on the actual module noise. The noise also depends on the hybrid
temperature increasing by \mbox{$\rm\sim\!6~{\rm electrons}$} per degree
Celsius. Since under standard conditions at the LHC the modules will operate
with a thermistor temperature near $\rm-7~^{\circ}C$ ---whilst the reported
measurements took place at \mbox{$\rm\sim10~^{\circ}C$}---, a lower noise level
than the one obtained during quality control tests is expected during running.
Furthermore, the implementation of the grounding, shielding and power
distribution scheme guarantee the stability and low-noise operation of the
modules when assembled into large structures (disks, cylinders), as
demonstrated in subsequent noise measurements \cite{integration}.

\begin{figure}[ht]
\begin{minipage}[c]{0.48\linewidth}
  \centering\epsfig{file=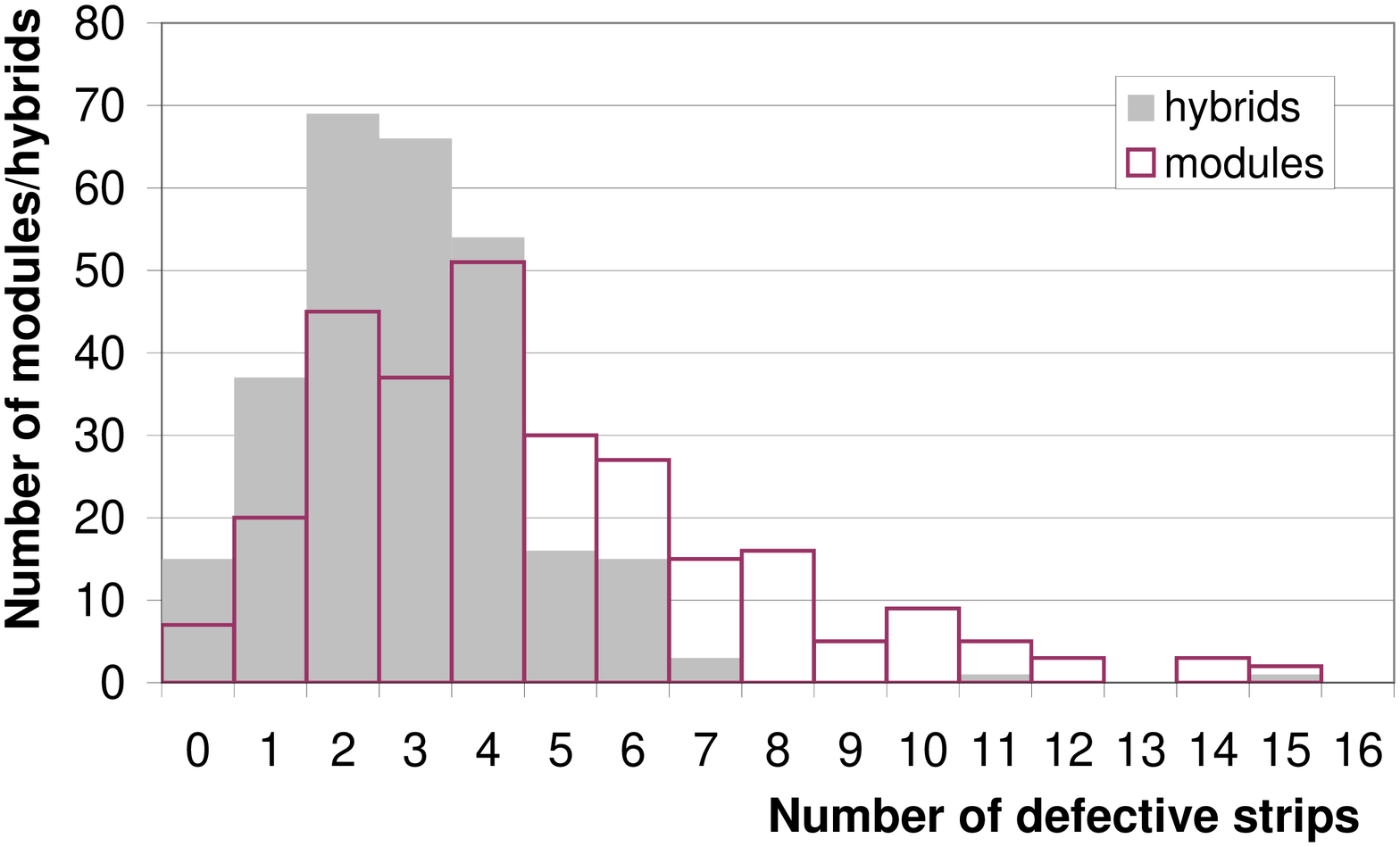,width=\linewidth,clip=}
\caption{Number of defective strips distribution per hybrid and per module.}
\label{fig:defects1}
\end{minipage}\hfill
\begin{minipage}[c]{0.48\linewidth}
  \centering\epsfig{file=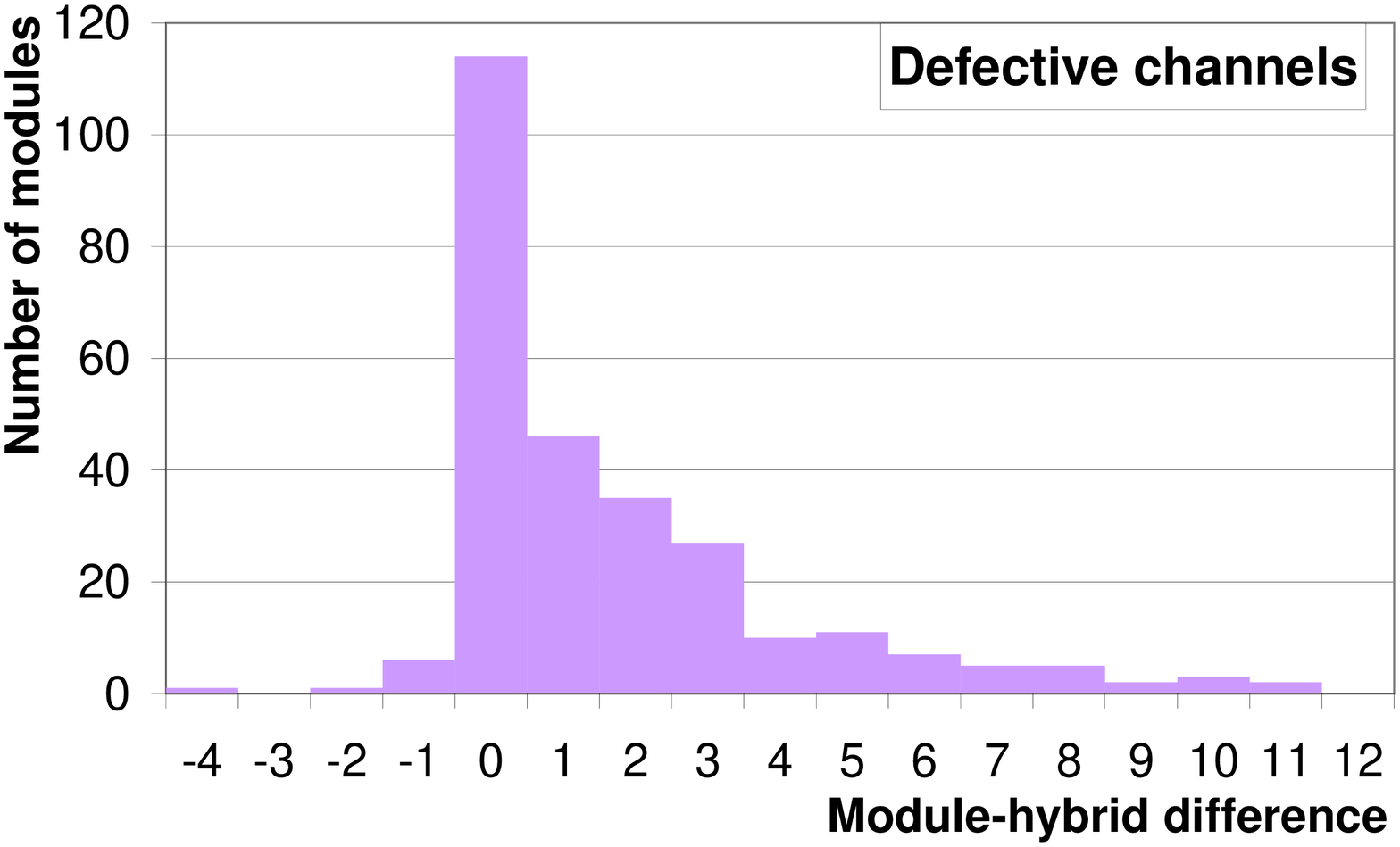,width=\linewidth,clip=}
\caption{Distribution of number of additional defective strips introduced
during assembly.} \label{fig:defects2}
\end{minipage}
\end{figure}

The distributions of the number of defective channels per unassembled hybrid
and per module are shown in Fig.~\ref{fig:defects1}. These include types of
critical defects, such as dead, stuck, noisy channels, channels that have not
been wire-bonded to the strips and channels that cannot be trimmed
(\emph{untrimmable}). These channels are masked during normal operation and are
practically \emph{lost}. Also included in the distributions are less critical
faults such as low or high gain (or offset) with respect to the chip average,
which are still operational.

In Fig.~\ref{fig:defects2} the number of faulty channels induced during
assembly is presented, i.e.\ those not present during the hybrid QC. These are
mostly due to sensor defects, such as oxide pinholes, strip metal shorts or
opens (sometimes caused by scratches) and to a lesser extend to partly-bonded
or un-bonded channels. The negative values represent trimmable channels tagged
as untrimmable during hybrid testing.

\begin{figure}[ht]
\begin{minipage}[c]{0.48\linewidth}
  \centering\epsfig{file=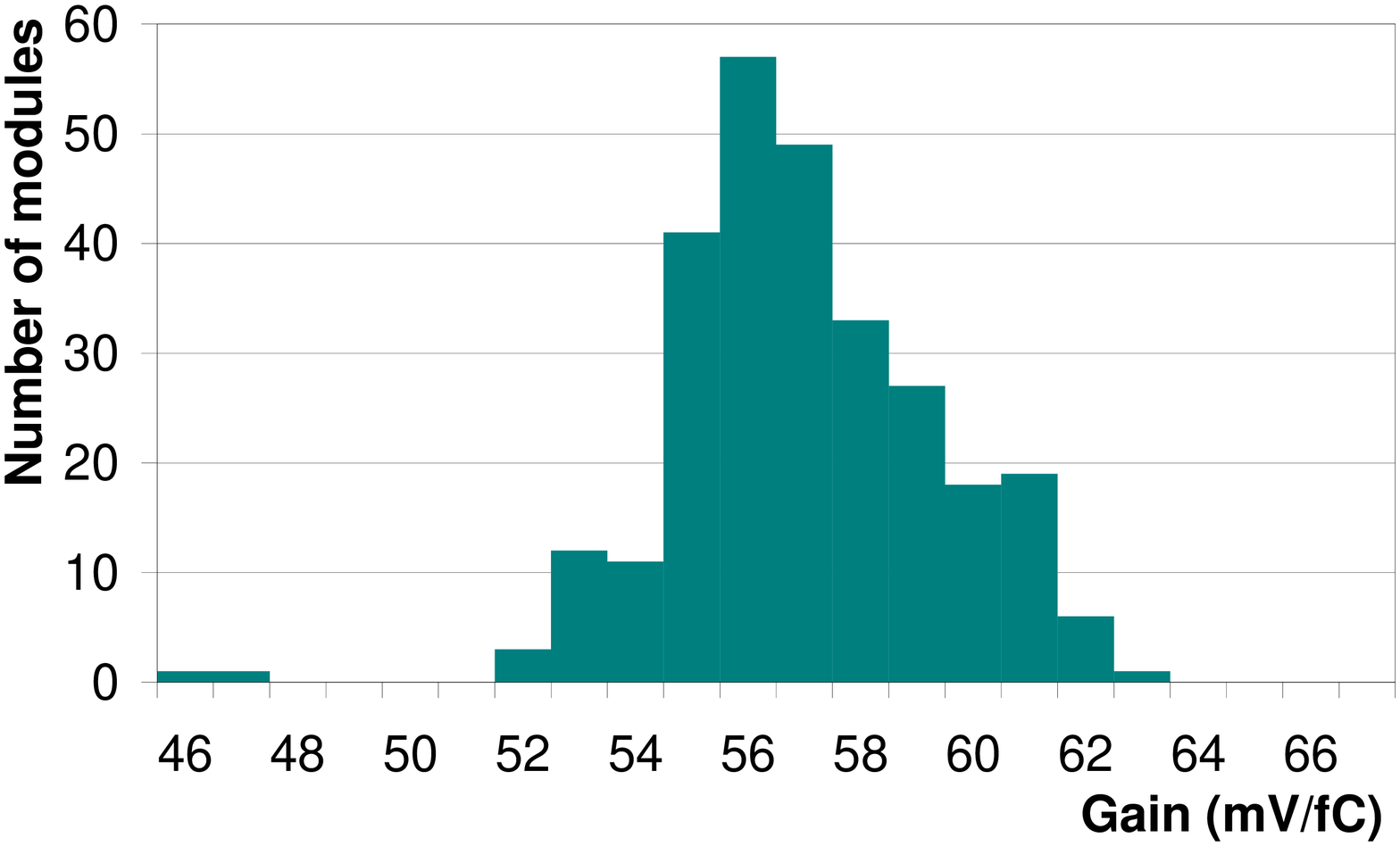,width=\linewidth,clip=}
  \caption{Average per module gain distribution for a discriminator threshold of 2~fC.} \label{fig:gain}
\end{minipage}\hfill
\begin{minipage}[c]{0.48\linewidth}
  \centering\epsfig{file=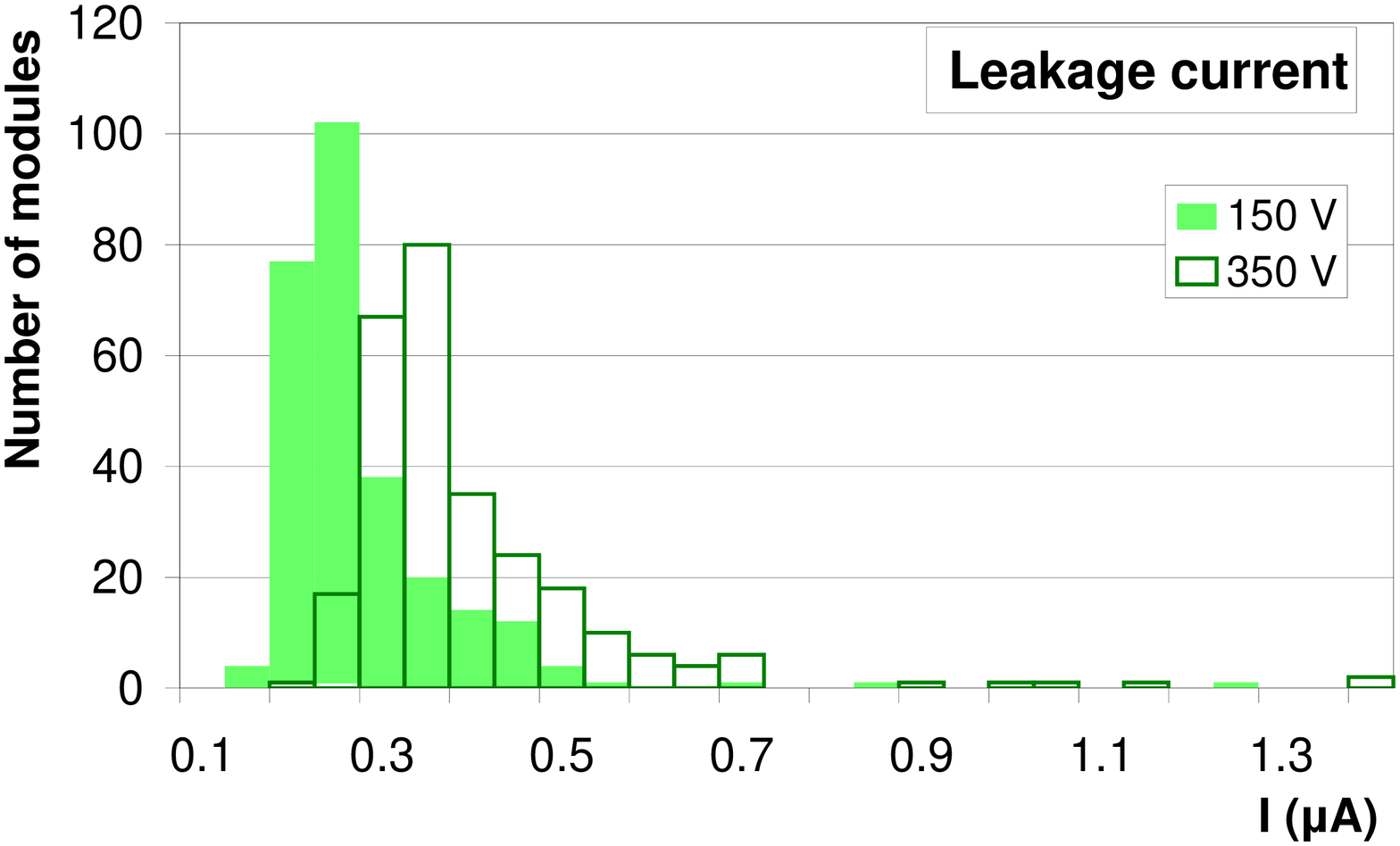,width=\linewidth,clip=}
  \caption{Leakage current distributions at $\rm20~^{\circ}C$ for a bias voltage of 150~V and 350~V.}\label{fig:iv}
\end{minipage}
\end{figure}

In Fig.~\ref{fig:gain} the average gain per module is shown for all qualified
forward modules. The average gain value is about 57~mV/fC with an RMS of
2.3~mV/fC at a discriminator threshold of 2~fC and it is of the same level as
the one obtained from system tests.

The distributions of leakage current for a bias voltage of 150~V and 350~V are
shown in Fig.~\ref{fig:iv}. These results, taken at a hybrid temperature of
\mbox{$\rm\sim\!20~^{\circ}C$}, represent the final current measurements during
module assembly and they do not include modules rejected because of high
current. The values spread partly reflects the strong dependence of the leakage
current on temperature, roughly doubling every $\rm7~^{\circ}C$. The last
11~modules were assembled with sensors demonstrating relatively high leakage
current that were designated for use during the (pre-)qualification phase.
After training at progressively increasing high voltage for long periods of
time (1--2 days), all of these modules, apart from one (\#160), were
successfully recovered delivering an acceptable $IV$ curve.

\subsection{Defective chips}\label{sub:elec_defects}

Some exceptional ASIC cases and the remedies applied are discussed in the
following subsections. Apart from them, no other serious condition, as far as
the module front-end electronics is concerned, was encountered and none of the
modules was rejected because of poor electrical performance.

\paragraph{Large gain spread:}

\begin{figure}[ht]
 \centering
 \subfloat[$I_{\rm sh}=20~{\rm\mu A}$]{
    \label{fig:M67:a}
    \epsfig{file=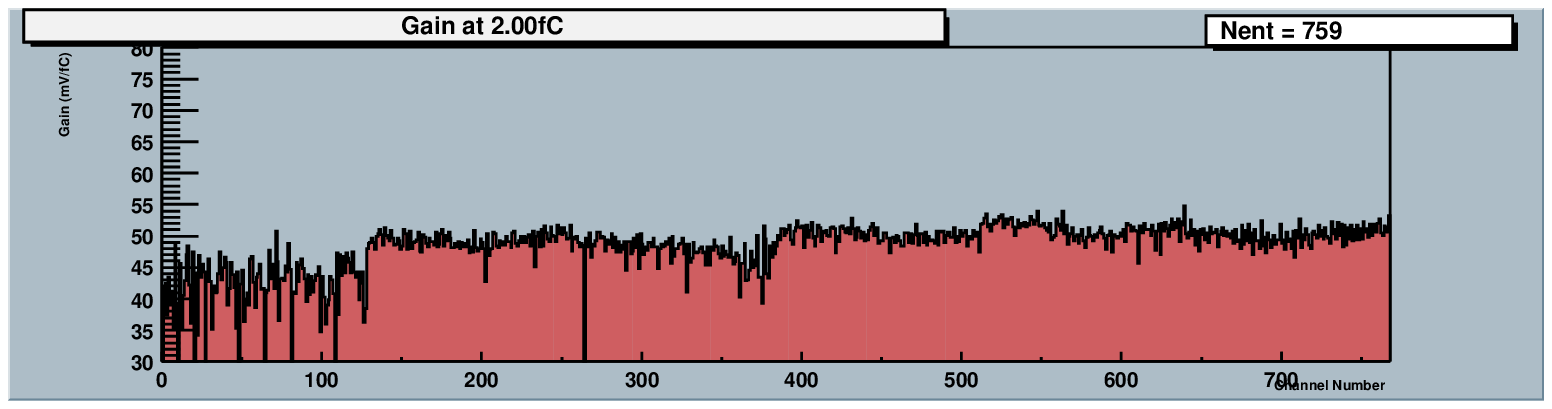,angle=-90,width=0.3\linewidth,clip=}
 }\hspace{0.05\linewidth} 
 \subfloat[$I_{\rm sh}=15~{\rm\mu A}$]{
    \label{fig:M67:b}
    \epsfig{file=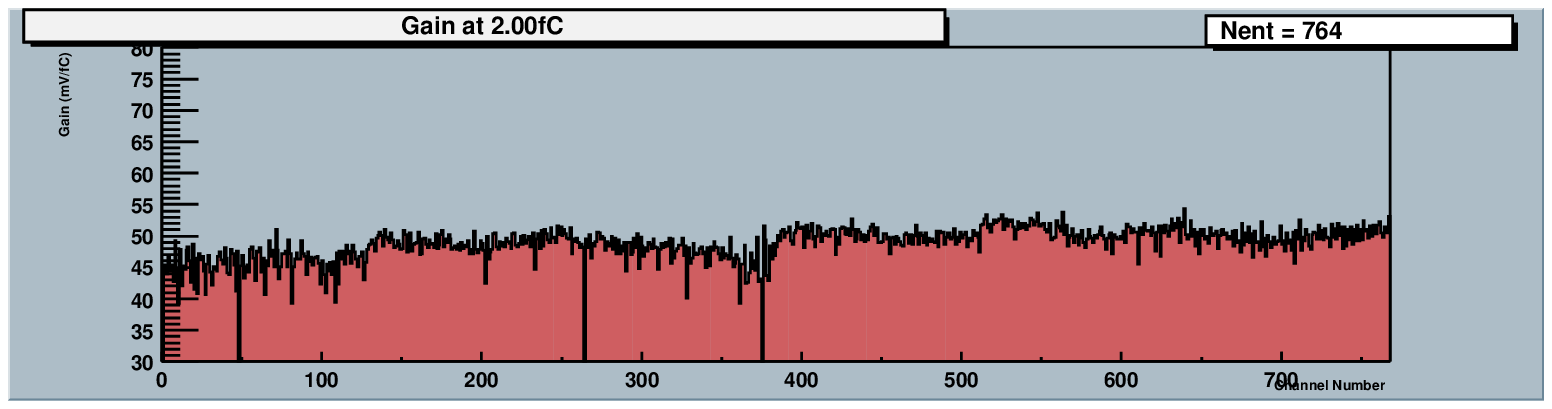,angle=-90,width=0.3\linewidth,clip=}
 }
\caption{Gain vs.\ channel number for the front side (only the first two ASICs
are shown) of Valencia module \#67; M0 chip shows LGS effects.} \label{fig:M67}
\end{figure}

A particularly low and Large Gain Spread (LGS) was observed in the M0 chip of
Valencia module \#67\footnote{The modules assembled in Valencia are numbered as
2022039000XXXX, where XXXX ranges from \#0004 to \#0160 for the middle modules
and \#1001 to \#1125 for the outer ones. Modules \#1 to \#3 were assembled
during the pre-qualification period and they are not considered as production
modules.} as shown in Fig.~\ref{fig:M67}\subref{fig:M67:a}. Such occurrences,
which are due to the sensitivity of individual chips to small variations of the
operating conditions, are known and they are treated by lowering the shaper
current, $I_{\rm sh}$ \cite{ciocio}. The LGS chip M0 is recovered when the
shaper current is lowered from the nominal value of $I_{\rm sh}=20~{\rm\mu A}$
(cf.\ Fig.~\ref{fig:M67:a}) to $I_{\rm sh}=15~{\rm\mu A}$ (cf.\
Fig.~\ref{fig:M67:b}).

\paragraph{Reworked bonds on chips:}

\begin{figure}[ht]
\centering
\begin{minipage}[c]{0.32\linewidth}
  \centering\epsfig{file=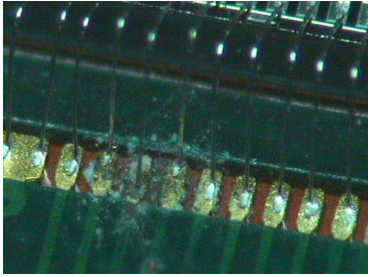,width=\linewidth,clip=}
\end{minipage} \hspace{0.05\linewidth} 
\begin{minipage}[c]{0.35\linewidth}
  \centering
\caption{Pad contamination in chip S1 of module \#1086, affecting the bonds
quality.} \label{fig:contamin}
\end{minipage}
\end{figure}

During the hybrid acceptance test, in one hybrid (20220554115113) no response
was received by five ASICs, namely S1--E5. Visual inspection showed that
contamination on pads of chip S1 caused three bond-wires to fail (see
Fig.~\ref{fig:contamin}). The hybrid was returned to Rutherford Appleton
Laboratory, reworked and passed successfully a warm characterisation test.
After being returned to Valencia, it was assembled into module \#1086 and
tested uneventfully.

In two other occasions, namely in modules \#76 and \#1078, bonds to detector
ground in ASICs (M0 and E5, respectively) were missing and were recovered
during wire bonding of the module.

\paragraph{Reworked high voltage contact:}

In one occasion, namely module \#88, the HV line was open-circuited, causing
all analog tests to fail in a consistent albeit unexpected manner. The problem
was readily confirmed using a multimeter and the electrical contact between the
HV finger of the hybrid and the sensor was restored.

\section{Bonding of CiS modules}\label{sec:CiS}

Apart from the standard modules assembly and testing, some additional studies
were performed in Valencia, aiming to address bonding issues that arose in
other assembly sites. To this effect, 11~middle modules assembled at the Max
Planck Institute (MPI) of Munich and one outer module from the University of
Melbourne were wire-bonded and tested. These modules were built with sensors
manufactured by CiS,\footnote{CiS Institut f\"{u}r Mikrosensorik gGmbH,
Konrad-Zuse-Stra{\ss}e 14, D-99099 Erfurt, Germany.} which differ from
Hamamatsu sensors with respect to leakage current and defective strips
\cite{endcap}.

During wire-bonding, some parameters had to be adjusted and some reworking was
necessary in order to achieve rigid wire connections. For instance, one of the
modules (L084) had poor metallisation on fan-in and efficient bonding was only
possible by reducing the bond speed. In other cases, glue had to be injected in
the central fan-in region. Furthermore, the bond rigidity was examined by
performing a pull test on a CiS dummy module, i.e.\ without ASICs, which had
been bonded at IFIC. The average pull strength necessary to remove the wire was
found to be $\rm13.8~g$, i.e.\ well above the lower limit of $\rm6~g$,
confirming the bonding firmness.

After bonding, five MPI modules and the Melbourne one were submitted to a
specific test aiming at identifying strips with oxide punch-through induced
during module wire-bonding. In these strips, the damaged bond-pads result in a
short between the p-implant and the aluminum pad, suspending thus the
AC-coupling between the strips and the readout electronics.

The identification of punch-through strips is carried out by applying the
following procedure: the detectors are biased through the hybrid at $\rm1~V$ in
a thermistor temperature of \mbox{$\rm\sim20~^{\circ}C$}; they are exposed to
light so that a leakage current of $\rm10~mA$ is acquired; and a three-point
gain test is finally performed. A punch-through strip appears subsequently in
the SCTDAQ results as a low-gain channel, however the outcome highly depends on
the testing conditions.

\begin{figure}[ht]
    \centering\epsfig{file=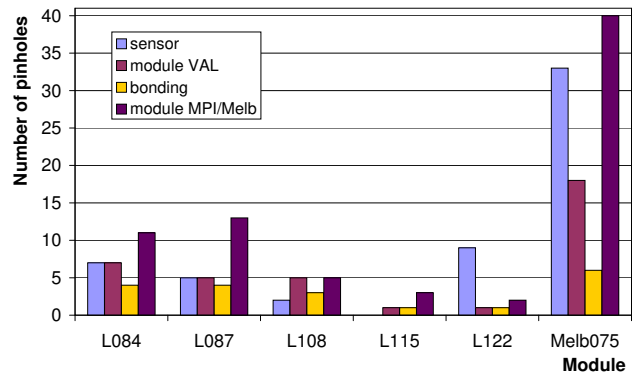,width=0.55\linewidth}
\caption{The number of pinholes (punch-through strips) per module observed in
various measurements.} \label{fig:pinholes}
\end{figure}

The pinhole results per module are summarised in Fig.~\ref{fig:pinholes} for
each measurement performed: obtained during sensor QA (`sensor'), observed in
Valencia (`module VAL'), induced by bonding (`bonding') and finally detected at
the assembly site (`module MPI/Melb'). Due to the different setup, testing
procedure and conditions, the number of pinholes detected varies between
different sites. For instance, some pinholes detected in sensors were not
observed in Valencia after bonding, but they were subsequently found at MPI or
Melbourne. The same results were acquired at IFIC for the Melbourne module,
when the test was repeated with an increased current value of $\rm20~mA$, i.e.\
with more light. The number of strips with punch-through induced by bonding
(3.2 on average) is compatible with the respective induced in other CiS
assembling sites (e.g.\ MPI) and seems to be correlated with the number of
intrinsic pinholes present in the sensor.

\section{Overall results}\label{sec:results}

A total of 125~outer production modules and 157~long middle ones have been
successfully built and tested in Valencia, not taking into account the various
modules manufactured during the site pre-qualification period. The detailed
results of the quality control tests for all these modules have been uploaded
to the SCT production database \cite{DB}.\footnote{Available in the web site:
\href{http://ific.uv.es/sct/modules/}{http://ific.uv.es/sct/modules/}}

\begin{figure}[ht]
    \centering\epsfig{file=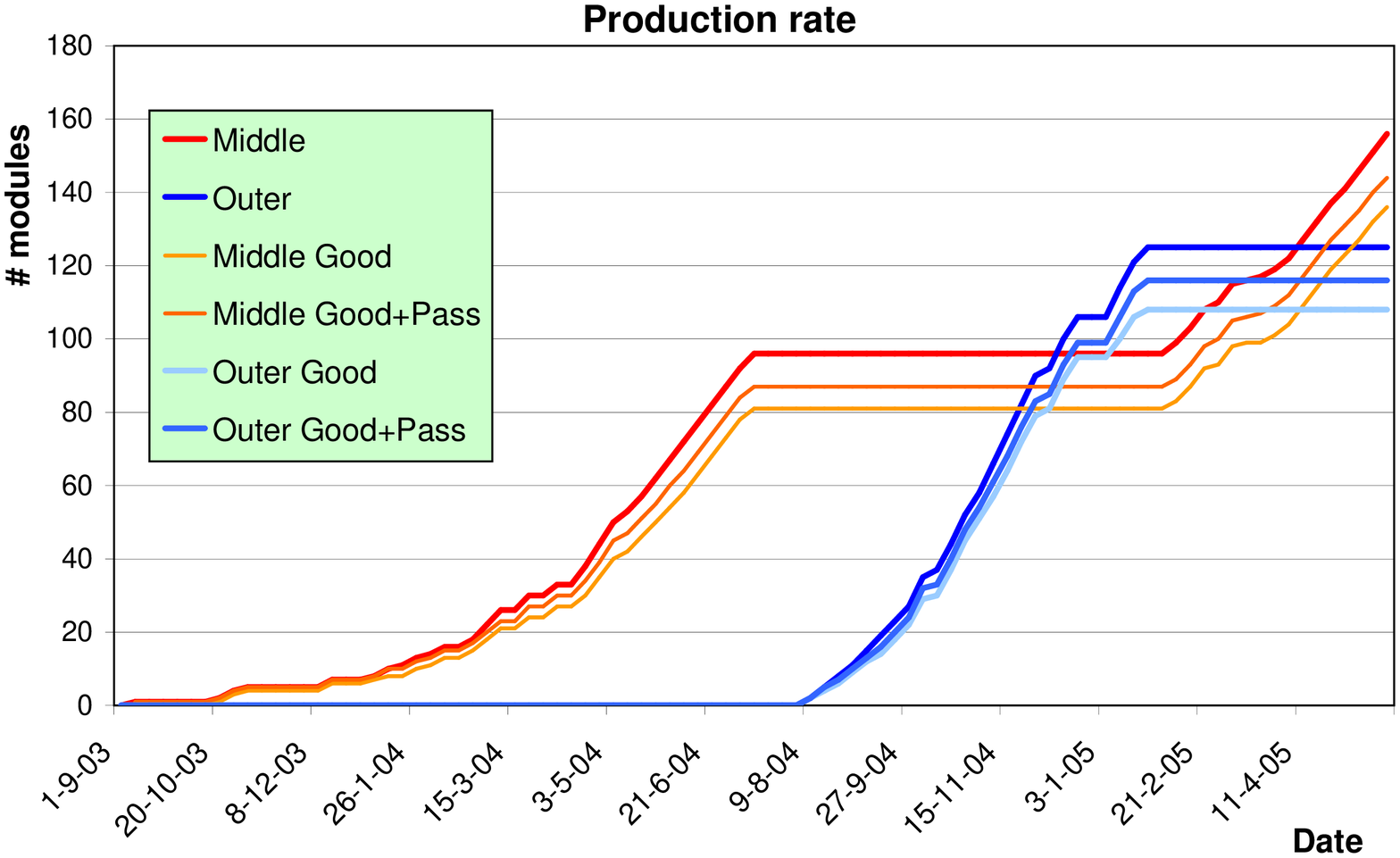,width=0.6\linewidth}
    \caption{Accumulated number of produced modules as a function of time.}
    \label{fig:rate}
\end{figure}

The module production took place from September~2003 till June~2005 at a rate
of five or eight modules per week as shown in Fig.~\ref{fig:rate}. For the long
middle modules only one set of assembly jigs produced mechanically stable
modules, therefore only one module per day was constructed, i.e.\ five modules
per week. The outer modules, on the other hand, were manufactured with two sets
of jigs, allowing the assembly of two modules per day. Nevertheless, the
time-consuming mechanical survey limited the rate to eight modules per week.
The production started with the building of the middle modules, continued with
those of the outer type, and finished with the extra middle ones. Before
switching to another type of modules, some dummy modules were built to insure
the correct configuration of the assembly and testing setups. The low rate
periods of correspond mostly to vacations or to lack of hybrids (just after
production startup). The few cases of device failure (wire-bonding machine,
metrology stand) were dealt with successfully and did not affect the production
rate.

\subsection{Geometry}\label{sub:geo_res}

\begin{figure}[ht]
\centering
\begin{minipage}[c]{0.48\linewidth}
  \centering\epsfig{file=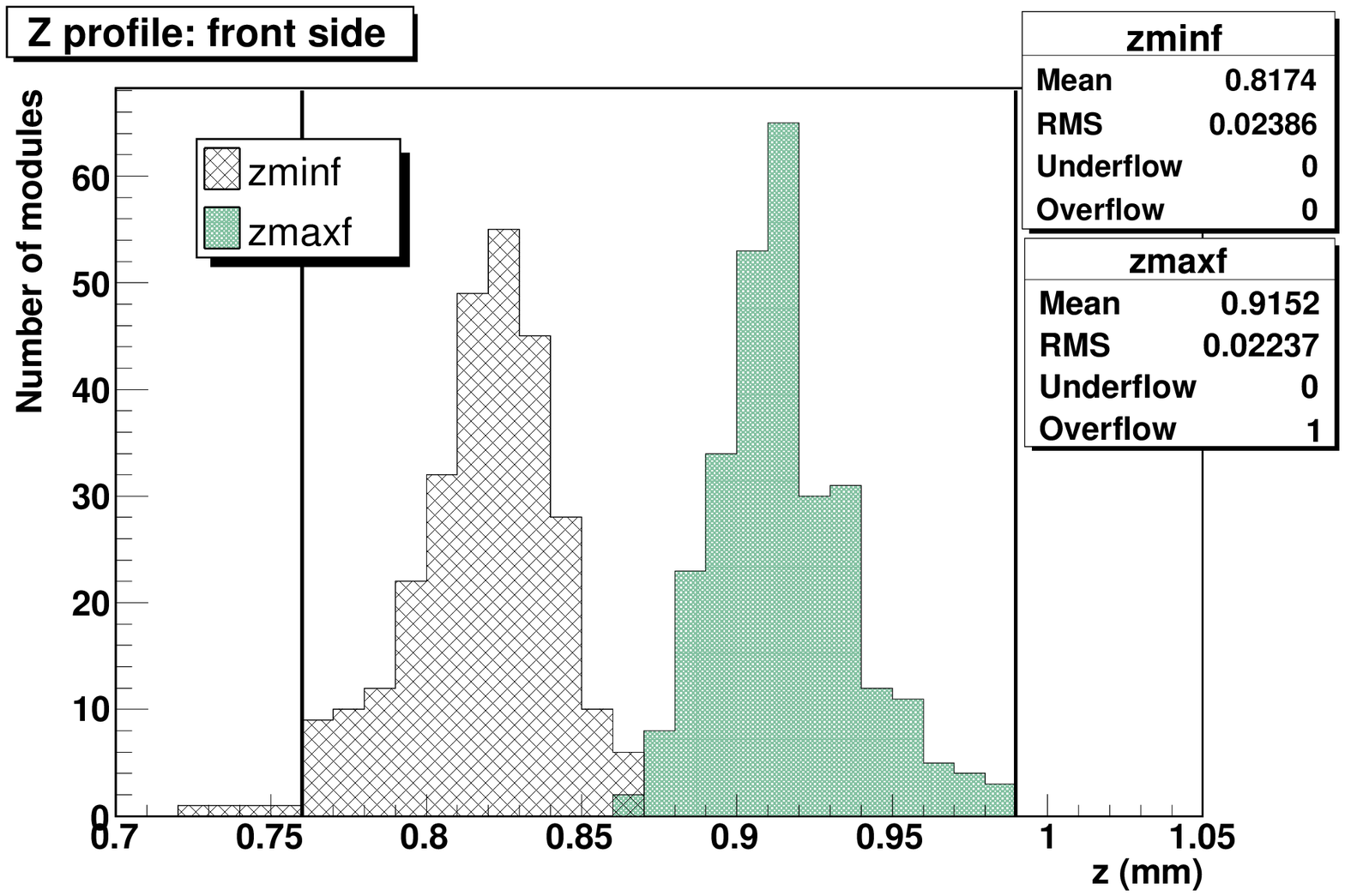,width=\linewidth}
\end{minipage} \hspace{0.02\linewidth} 
\begin{minipage}[c]{0.48\linewidth}
  \centering\epsfig{file=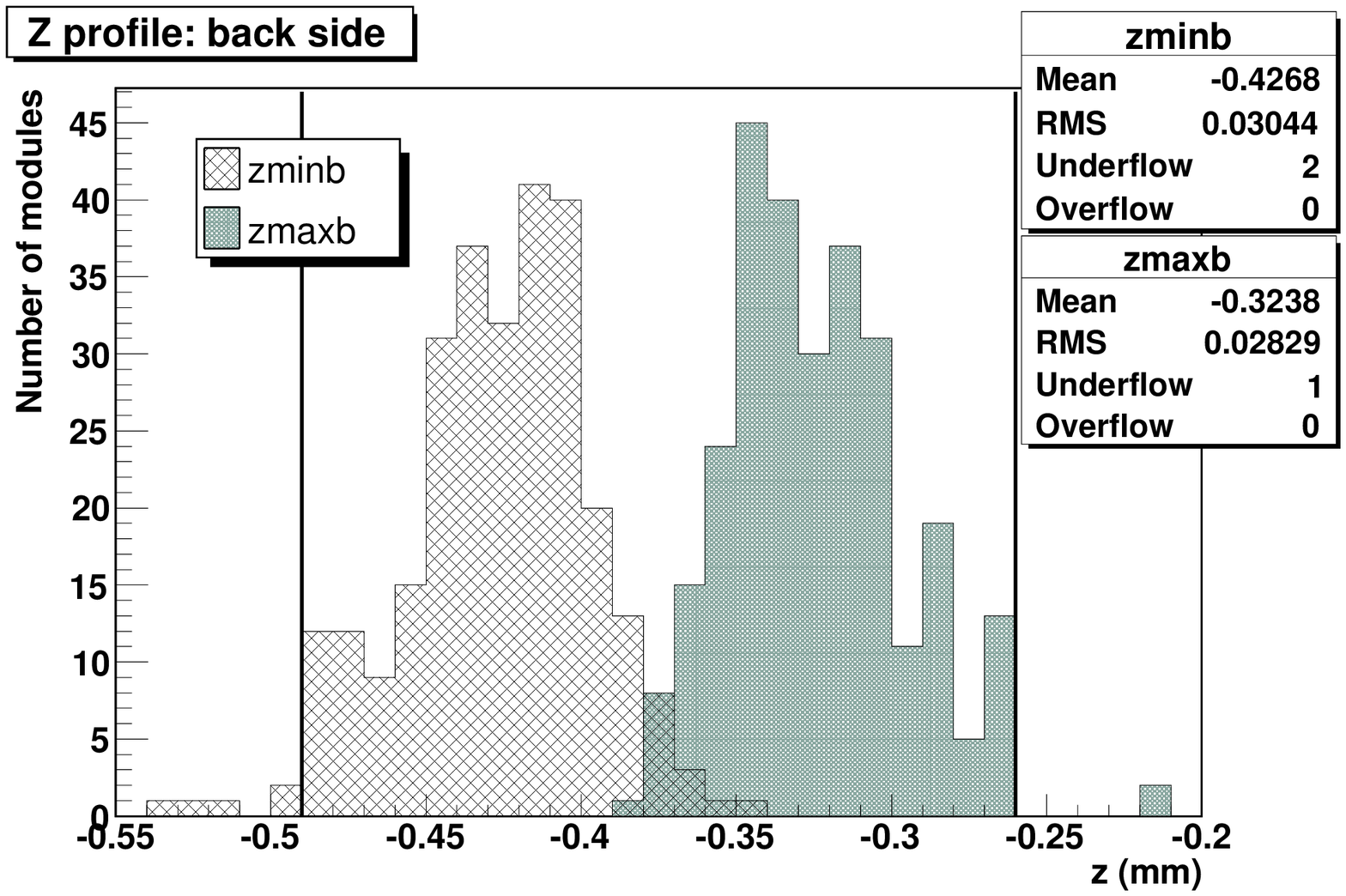,width=\linewidth}
\end{minipage}
\caption{Distributions of minimum-$z$ and maximum-$z$ for the front (left) and
back (right) sides. The black vertical lines designate the specified
tolerances. The underflows and overflows correspond to rejected
modules.}\label{fig:z}
\end{figure}

The geometrical parameters of the produced modules are summarised in
Table~\ref{tab:XY} for the $XY$ parameters and in Fig.~\ref{fig:z} for the $Z$
profile. In general, the $XY$ metrology results are clearly well within the
mechanical tolerances, and only $midyf$ shown a wider distribution, as
discussed in Sec.~\ref{sub:metro_res}. The $Z$ profile of the modules, on the
other hand, deviates from the specified dimensions only in a handful of cases,
in which however the module is still usable, as shown in the module
classification that follows.

\begin{table}[ht]
\begin{center}
\begin{tabular}{ l c c | l c c | l c c }                 \hline
  Parameter\T\B & Mean & RMS & Parameter & Mean & RMS & Parameter & Mean & RMS  \\
  \hline
  $a1$\T\B & \phm0.08 & 0.17 & $mhx$ &    -0.03 & 0.26 & $midxf$ & -0.08 & 0.17 \\
  $a2$     & \phm0.05 & 0.19 & $mhy$ & \phm0.01 & 0.29 & $midyf$ & -0.03 & 0.51 \\
  $a3$     & -0.04    & 0.15 & $msx$ & \phm0.07 & 0.40 & $sepf$  & -0.16 & 0.23 \\
  $a4$     & -0.02    & 0.20 & $msy$ & \phm0.11 & 0.36 & $sepb$  & -0.16 & 0.22 \\
  $stereo$ & -0.04    & 0.24 &       &          &      &         &       & \\ \hline
\end{tabular}
\caption{Summary of $XY$ parameters expressed as (\emph{value -- nominal}) $/$
\emph{tolerance} for the IFIC modules.} \label{tab:XY}
\end{center}
\end{table}

The module quality, as far as metrology is concerned, is similar to the one
reached in the other assembly sites of end-cap \cite{endcap} and barrel modules
\cite{barrel}. It is also comparable to that achieved in microstrip detector
modules of other experiments of the LHC ---which share the same stringent
requirements as ATLAS--- such as CMS \cite{CMS}.

\subsection{Electronics}

A resume of the production performance of the Valencia modules in terms of
readout is provided in Table~\ref{tab:elec}. The electrical properties acquired
are compatible with the ones collectively achieved by all forward module
assembly sites \cite{RT2005}. On average, two channels per module are lost,
i.e.\ unusable, representing 0.1\% of the total. The noise occupancy was also
kept well below the specified value throughout the module production.

\begin{table}[ht]
\begin{center}
\begin{tabular}{ l l c c }                 \hline
  & Module type\T\B & Mean & RMS \\ \hline
  \multirow{2}{*}{ENC (e$^-$)}\T\B & Middle & 1436 & 76 \\
                               & Outer  & 1547 & 68 \\
  \multirow{2}{*}{Noise occupancy} & Middle & $1.3\times10^{-5}$ & $1.5\times10^{-5}$ \\
                                   & Outer  & $2.3\times10^{-5}$ & $2.2\times10^{-5}$ \\
  Faulty channels         & All & 4.7\pho & 3.3\pho \\
  Lost channels           & All & 1.8\pho & 2.3\pho \\
  $I_{\rm leak}$ at 150~V & All & 0.32    & 0.19    \\
  $I_{\rm leak}$ at 350~V & All & 0.45    & 0.31    \\ \hline
\end{tabular}
\caption{Electrical properties of Valencia modules.} \label{tab:elec}
\end{center}
\end{table}

\subsection{Final yield}\label{sub:yield}

The last step in the module assembly is the final evaluation, in which the
outcome of all QC tests is taken into account. The modules are classified as:
\begin{compactdesc}
  \item[Good:]
if all mechanical and electrical parameters are within the specifications.
  \item[Pass:]
if the metrology measurements are within 15\% tolerance and a smooth
$IV$-curve, not exceeding the current limit of 80~$\rm\mu A$ is obtained up to
a minimum breakdown voltage of 350~V.
  \item[Hold:]
if electrical specifications are not met but at least a smooth IV curve is
obtained up to 350~V and all chips are responding. It may be usable if
reworked.
  \item[Fail:]
if the module does not match any of the above categories and cannot be
reworked.
\end{compactdesc}

\begin{figure}[ht]
    \centering\epsfig{file=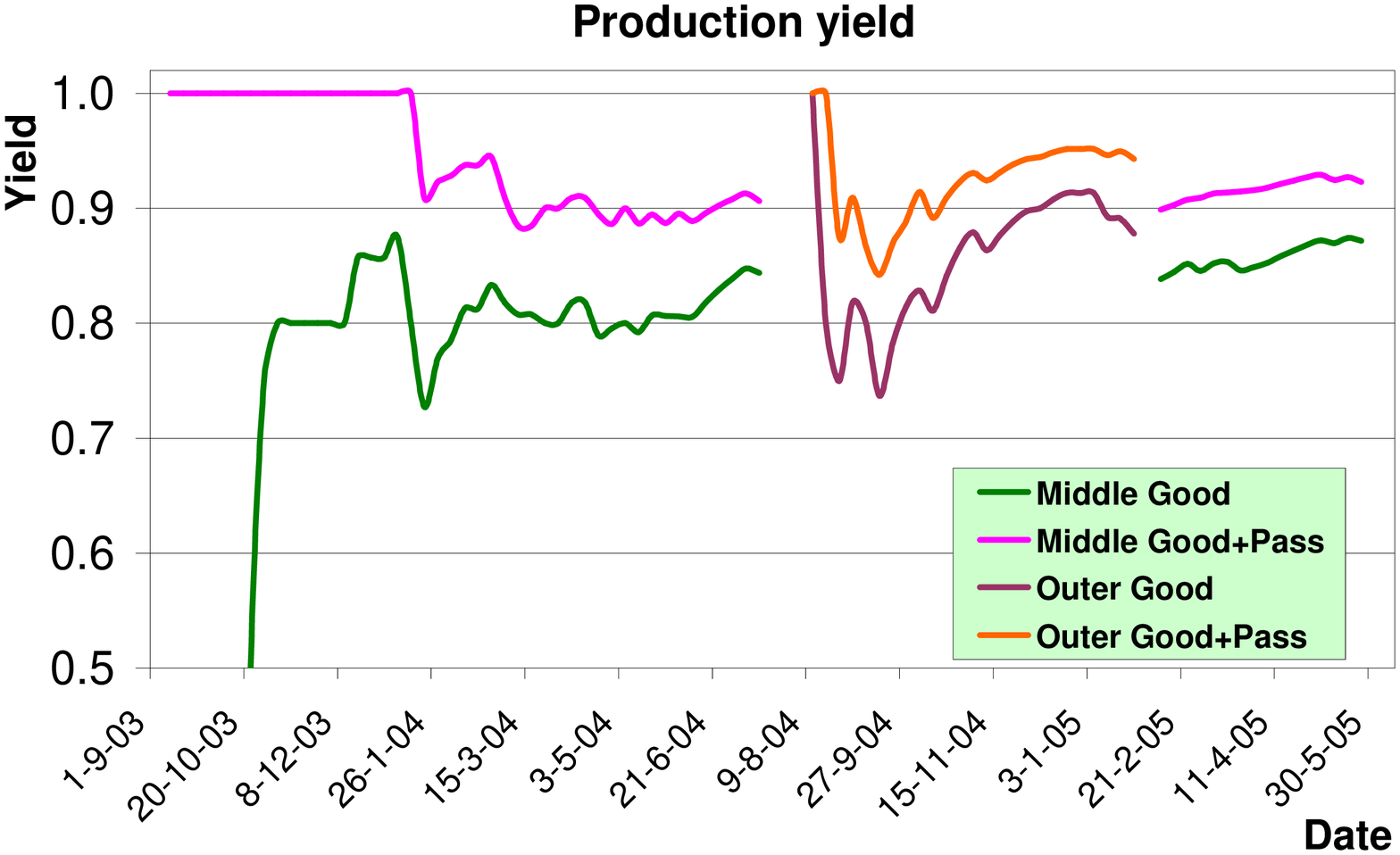,width=0.7\linewidth}
    \caption{The production yield as evolves with time (in weeks).} \label{fig:yield}
\end{figure}

The evolution of the yield, defined over the total number of built modules, is
drawn in Fig.~\ref{fig:yield} for the modules belonging in the \emph{Good\/}
and in the \emph{Good+Pass\/} category. It is clear from this graph that
although the production of both middle and outer modules started with a rather
low yield, as experience was accumulated, the yield increased.

The classification of the modules produced in Valencia according to the
aforementioned quality scheme is summarised in Table~\ref{tab:yield}. The yield
of the extra middle production is indeed significantly better that the initial
one; 91.8\% vs.\ 84.4\% for the \emph{Good\/} category as observed in
Fig.~\ref{fig:yield}.

\begin{table}[ht]
\begin{center}
\begin{tabular}{ l r r r r r }                 \hline
  Type \T\B        & Good & Good+Pass & Hold & Fail & Total \\ \hline
  Outer \T\B       & 109 (87.2\%) & 118 (95.4\%) & 6 (4.8\%) & 1 (0.8\%) & 125 \\
  Middle allocated & 81 (84.4\%)  & 87 (90.6\%)  & 7 (7.3\%) & 2 (2.1\%) & 96 \\
  Middle extra     & 56 (91.8\%)  & 58 (95.1\%)  & 2 (3.3\%) & 1 (1.6\%) & 61 \\
  Middle total     & 137 (87.3\%) & 145 (92.4\%) & 9 (5.7\%) & 3 (1.9\%) & 157 \\ \hline
  All \T\B         & 246 (87.2\%) & 263 (93.3\%) & 15 (5.3\%) & 4 (1.4\%) & 282 \\ \hline
\end{tabular}
\caption{Final yield of the Valencia modules.} \label{tab:yield}
\end{center}
\end{table}

\begin{figure}[ht]
 \centering
 \subfloat[\#27: fractured sensor.]{
    \label{fig:M27}
    \epsfig{file=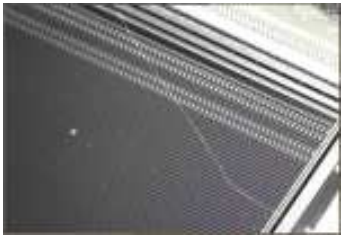,width=0.3\linewidth,clip=}
 }\hspace{0.05\linewidth} 
 \subfloat[\#39: scratched sensor.]{
    \label{fig:M39}
    \epsfig{file=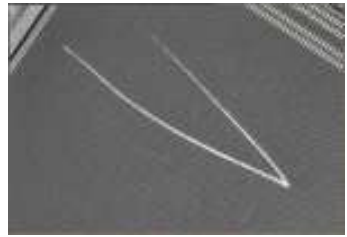,width=0.3\linewidth,clip=}
 }\\
 \subfloat[\#71: scratched fan-in.]{
    \label{fig:M71}
    \epsfig{file=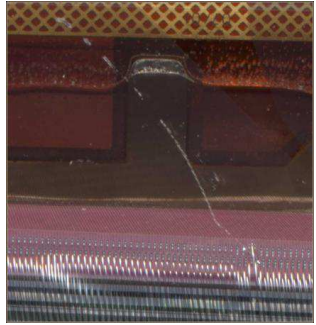,width=0.27\linewidth,clip=}
 }\hspace{0.05\linewidth} 
 \subfloat[\#34: broken ceramic in spine.]{
    \label{fig:M34}
    \epsfig{file=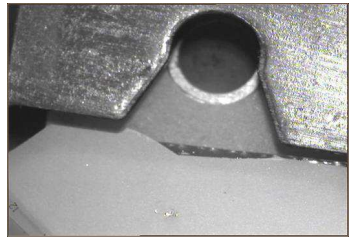,width=0.3\linewidth,clip=}
 }\caption{Cases of accidental damage to modules.} \label{fig:broken}
\end{figure}

Only four modules out of the 282 (1.4\%) fall into the \emph{Fail\/} category,
i.e.\ they are not suitable for operation within the ATLAS detector. These are
the following modules:
\begin{compactdesc}
  \item[\#24] Middle module in which a fan-in set of the wrong type was glued.
  \item[\#27] Middle module with a broken sensor, shown in Fig.~\ref{fig:broken}.
  \item[\#100] Middle module with damaged wire bonds and one chip (S11) dead after
   an accident during manipulation.
  \item[\#1006] Outer module with a fractured sensor.
\end{compactdesc}

The 15~modules belonging to the \emph{Hold\/} class, i.e.\ those which may be
repaired, amount to 5.3\% of the total number of assembled modules. These cases
were due to one or the combination of some of the following reasons:
\begin{compactdesc}
  \item[Metrology.]
Seven occurrences of failure in the $Z$-profile and three in the $XY$ (two with
$midyf$ and one with $a4$ out-of-spec).
  \item[Scratches.]
Two cases in sensors, two in fan-ins and one in spine ceramic. This is mostly
due to accidents while handling the components or the assembled modules. In
some cases the wire bonds were also affected resulting in failed electrical
tests (many unbonded or low-gain channels).
  \item[Fractures.]
Two cases of broken spines in the ceramic part near the hole, which is very
sensitive.
  \item[Leakage current.]
Three cases occurred; the one was due to oil being spilled on the module while
in the electrical tests box through the dry air supply. The tubes were replaced
and the setup was thoroughly cleaned before resuming the tests.
\end{compactdesc}

The majority of the problematic modules (\emph{Fail\/} or \emph{Hold\/}) was
due either to manipulation accidents or to the mechanical parameters not
conforming to the stringent specifications set by the collaboration.

\section{Conclusions}\label{sec:conclusions}

A series of assembling, bonding and testing devices have been developed, built
and installed at the silicon detectors laboratory at IFIC Valencia for the
construction of ATLAS SCT silicon detector modules. The assembly and quality
control of the 12\% of the forward ATLAS SCT modules has been successfully
completed at this site. In total 282~outer and long-middle modules have been
constructed during an 18-months period. The overall yield of the modules with
acceptable mechanical and electrical performance is 93.3\%, amounting to
118~outer and 145~inner modules. The electrical functionality of the modules
was repeatedly tested during the disk macro-assembly and after the disks were
put together into end-caps \cite{integration}, demonstrating equally
satisfactory performance.

\acknowledgments

This work has been performed within the ATLAS SCT Collaboration, and we thank
collaboration members for helpful discussions before and during the execution
of the production. We have made use of assembly techniques, software packages
and tooling components which are the result of collaboration-wide efforts.

We acknowledge support of CICYT (Spain) under the project FPA2003-03878-C02-01
and of the EU under the RTN contract: HPRN-CT-2002-00292, ``The 3rd Generation
as a Probe for New Physics: Technological and Experimental Approach.''


\end{document}